\documentclass[a4paper,11pt]{article}

\usepackage{amsmath}
\usepackage{graphicx}
\usepackage{amsfonts}
\usepackage{amssymb}

\newtheorem{theorem}{Theorem}[section]
\newtheorem{rema}{Remark}[section]
\newtheorem{assump}{Assumption}[section]
\newtheorem{propo}[rema]{Proposition}
\newtheorem{defi}[rema]{Definition}

\newtheorem{corol}[rema]{Corollary}

\newcommand{\sect}[1]{\setcounter{equation}{0}\section{#1}}

\newcommand{\bc}{\begin{center}}
\newcommand{\ec}{\end{center}}
\def\ba#1{\begin{array}{#1}\displaystyle}
\newcommand{\ea}{\end{array}}

\newcommand{\beq}{\begin{equation}}
\newcommand{\eeq}{\end{equation}}
\newcommand{\beqa}{\begin{eqnarray}}
\newcommand{\eeqa}{\end{eqnarray}}
\newcommand{\no}{\nonumber}
\newcommand{\n}{\nonumber\\}
\newcommand{\bi}{\begin{itemize}}
\newcommand{\ei}{\end{itemize}}

\def\lt#1{\left#1}
\def\rt#1{\right#1}
\def\t#1{\tilde{#1}}

\def\b#1{\bar{#1}}
\def\frc#1#2{\frac{#1}{#2}}

\newcommand{\p}{\partial}

\newcommand{\bra}{\langle}
\newcommand{\ket}{\rangle}

\newcommand{\N}{{\mathbb{N}}}
\newcommand{\R}{{\mathbb{R}}}
\newcommand{\C}{{\mathbb{C}}}

\newcommand{\uD}{{\mathbb{D}}}

\newcommand{\ep}{\epsilon}
\newcommand{\varep}{\varepsilon}
\newcommand{\bs}{\setminus}

\newcommand{\id}{{\rm id}}

\newcommand{\halmos}{\rule{1ex}{1.4ex}}
\newcommand{\eproof}{\hspace*{\fill}\mbox{$\halmos$}}
\newcommand{\proof}{{\em Proof.\ }}

\newcommand{\conf}{{\cal S}}

\newcommand{\ev}{{\cal E}}
\newcommand{\evs}{{\cal V}}
\newcommand{\sgm}{\sigma}

\newcommand{\tou}{{\cal X}}
\newcommand{\toua}{{\cal Y}}
\newcommand{\sam}{{\cal M}}
\newcommand{\pres}{\rho}

\newcommand{\inter}{I}
\newcommand{\exter}{E}

\newcommand{\supp}{{\rm supp}}

\newcommand{\non}{{!}}

\newcommand{\blim}{\lim}

\def\cl#1{\overline{#1}}

\begin{document}

\begin{titlepage}

\begin{center}
{\Large {\bf
Conformal loop ensembles and the stress-energy tensor.

I. Fundamental notions of CLE} 

\vspace{1cm}

Benjamin Doyon}

Department of Mathematical Sciences, Durham University\\
South Road, Durham DH1 3LE, UK\\
email: benjamin.doyon@durham.ac.uk

\end{center}

\vspace{1cm}

\noindent This is the first part of a work aimed at constructing the stress-energy tensor of conformal field theory as a local ``object'' in conformal loop ensembles (CLE). This work lies in the wider context of re-constructing quantum field theory from mathematically well-defined ensembles of random objects. The goal of the present paper is two-fold. First, we provide an introduction to CLE, a mathematical theory for random loops in simply connected domains with properties of conformal invariance, developed recently by Sheffield and Werner. It is expected to be related to CFT models with central charges between 0 and 1 (including all minimal models). Second, we further develop the theory by deriving results that will be crucial for the construction of the stress-energy tensor. We introduce the notions of support and continuity for CLE events, about which we prove basic but important theorems. We then propose natural definitions of CLE probability functions on the Riemann sphere and on doubly connected domains. Under some natural assumptions, we prove conformal invariance and other non-trivial theorems related to these constructions. We only use the defining properties of CLE as well as some basic results about the CLE measure. Although this paper is guided by the construction of the stress-energy tensor, we believe that the theorems proved and techniques used are of interest in the wider context of CLE. The actual construction will be presented in the second part of this work.

\vfill

{\ }\hfill 26 May 2009

\end{titlepage}

\tableofcontents

\sect{Introduction}

Many-body systems subject to fluctuations (thermal or quantum) often exhibit collective behaviours that are hard to predict from the local, short-range interactions amongst individual components. These collective behaviours are at the basis of the most interesting and surprising properties of physical systems. They can be understood as the formation of ``new'' collective objects from large numbers of components acting together: for instance, clusters of components taking correlated values. Such objects are always more or less present, but there are situations where they dominate the large-distance physics. That is, in these situations, statistical fluctuations do not destroy these objects, yet give them nontrivial statistics, which can then be used to fully describe the many-body system at large distances. Naturally, it is hard to observe these collective objects in many-body statistical models, and to provide a description of their statistics. However, a clear sign of their presence is when correlations can be observed up to large distances, amongst components that are only connected through large chains of local interactions. This is observed, for instance, through singularities of response functions (divergence of correlation lengths). The leading characteristics of these correlations is their invariance, or covariance, under re-scaling. The physical idea behind this is that it would be unnatural to have a pre-determined large-distance scale, controlling collective objects, from the microscopic scales of local interactions. The presence of such large-distance, scale invariant correlations is what defines critical points (in the parameter space), where the system is said to be critical.

Critical points occur naturally when the system is on the verge of a second order phase transition, where a macroscopic number of components are to go from an ordered to a disordered state (or {\em vice versa}). For instance, in the context of magnetism, a critical point occurs at the Curie temperature, where the system goes from a phase where local magnetic moments are mostly aligned to one where they are not. Although magnetic moments interact only with their nearest neighbours, at the Curie temperature there are statistical correlations up to large distances. Indeed, at this point there is no preferred magnetic direction of alignment, but neither enough thermal disturbance to force disorder. Hence, the system reacts sensitively to external disturbances, however small: there are macroscopic effects following from local disturbances.

The physical theory describing collective behaviours of many-body systems, based on the properties of correlations observed near critical points, is quantum field theory (QFT). More precisely, QFT is a theory for exact predictions in the so-called scaling limit: the result of approaching a critical point while looking at larger and larger distances, in proportion to the growth of the correlation length. QFT is universal: one model applies to many microscopic models. This is because collective behaviours in general do not depend on the exact form of local interactions, but only on some global characteristics. Universality goes beyond many-body systems: certain collective behaviours described by QFT are what is understood in modern theoretical physics as the fundamental particles of high-energy physics. However, very few models of QFT can be proven to actually describe the result of the scaling limit of some microscopic model. In this sense, and in many others, QFT is a physical, and not a mathematical, theory.

Exactly critical models offer the hope of a better understanding. The scaling limit exactly at a critical point is obtained by keeping the system at criticality while we look at larger and larger distances. Critical QFT models are scale invariant, and in many cases, there is also invariance under Euclidean rotation and translation. Scale, rotation and translation invariance, along with locality of QFT, are expected to imply that the full group of space-transformation invariance is the conformal group. In two dimensions, this is a strong statement, since there is an infinite-dimensional space of local generators. The theory describing this is two-dimensional conformal field theory (CFT) (see, for instance, the book \cite{DFMS97}).

The algebraic approach to CFT is based on general QFT axioms along with expected properties of these local generators. There are various ways of mathematically developing the algebraic theory; one well-developed way is that of vertex operator algebras and their modules \cite{LL04}. Each element of this algebraic theory is expected to correspond to the scaling limit of an element in the underlying statistical model. For instance, elements of the vertex operator algebra are local fields associated to symmetries (local currents and their descendants, in the QFT terminology), and elements of the modules are other types of local fields (primary fields and their descendants)\footnote{The concept of local fields has a precise definition in the context of QFT; they can often, but not always, be seen as the scaling limit of statistical variables defined on a finite number of lattice sites in the neighbourhood of a point.}. The scaling limit of correlation functions of statistical variables has a meaning through tensor products of vertex operator algebra modules. The local fields associated to conformal transformations form the Virasoro vertex operator algebra, which is always at least a sub-algebra. This sub-algebra is arguably the most important part of the algebraic theory. The generating element of the Virasoro vertex operator algebra is called the stress-energy tensor. It is this element, and its relation to other local fields, that gives rise to many of the non-trivial predictions of CFT, and that makes it integrable.

However, this approach presents many problems. First, the relation with the underlying statistical variables is conjectural, as the construction is purely algebraic. Second, predictions are mostly restricted to models with nice algebraic properties (rational models). Third, a deeper problem has to do with the understanding of conformal invariance itself. In any given CFT model, there is in fact a very restricted set of conformal symmetry transformations. For instance, on the Riemann sphere, these are the M\"obius maps (global conformal transformations). However, local generators of other conformal transformations are also considered, and in the standard approach their properties are derived as if they were generating symmetry transformations, up to the conformal anomaly (leading to the central charge of the Virasoro algebra). Properties of the stress-energy tensor, in particular the conformal Ward identities, follow from these considerations. There are reasons for this, based on locality, but a clearer understanding would be useful.

The recent developments of Schramm-Loewner evolution (SLE) \cite{L23,S00} (for reviews, see \cite{KN04,C05}) and of conformal loop ensembles (CLE) \cite{W05a,W05b,Sh06,ShW07a,ShW07b} provide an entirely new viewpoint on CFT, describing the collective objects themselves instead of the scaling limit of local statistical variables. This suggests that a fuller understanding of CFT and a better connection with critical statistical models in their scaling limit can be obtained in two steps. First, we have to prove that these descriptions of collective objects indeed emerge from underlying statistical models. Second, we have to construct the algebraic CFT structure from these collective objects, where ideally the local fields are certain random variables of these objects, and correlation functions are mathematical expectations of products of these. We would like to see the second step as {\em constructive CFT}, where the CFT description of collective behaviours is explicitly deduced from the collective objects themselves -- an alternative to the usual constructive field theory ideas related to random distributions.

This paper is the first part of a work making progress on the second step: the explicit construction of the stress-energy tensor in terms of the random loops described by CLE (in the ``dilute'' regime). This provides a clear understanding of the origin of the conformal Ward identities, and identifies the central charge and the CFT partition function through the measure on these collective objects.

The actual construction of the stress-energy tensor, and a more extensive discussion of the QFT implications of this construction, will be presented in the second part of this work. In the present paper, we only discuss CLE. We provide an introduction to CLE, attempting to discuss it in a mathematically precise way that is accessible to theoretical physicists. We also give interpretations of the main CLE axioms, and discuss the connection to CFT. We then develop the theory further, obtaining results that will be essential in the construction of the stress-energy tensor. We tried to be precise both in the formulations and proofs of the results, since the construction of the second part of this work relies on somewhat subtle theorems. Our proofs are based only on the defining axioms of CLE as well as some of its basic properties, and some basic language of measure theory is used (see, for instance, \cite{H50}).

Two main notions are introduced: that of continuity and Lipschitz continuity of CLE events, and that of support. Continuity says that the probability of an event does not change much under small perturbations of the domain of definition. This ``functional analysis'' of events is not often discussed, although something similar is assumed of CFT correlation functions from the outset. On the other hand, the support of an event tells us where the event ``lies'', or in which regions CLE loops would affect its evaluation. This should be paralleled with the notion of locality of CFT fields. We show Lipschitz continuity for a family of events that will be at the heart of the stress-energy tensor construction (theorem \ref{theoslcev}), and we show continuity -- more precisely, a slightly stronger version than the usual continuity statement -- in general for events supported on domains (open sets) (theorem \ref{theocontevent}). The continuity statements that we establish are essential for the proofs of the main theorems in the present paper, because we do not make direct use of any construction of the CLE measure.

The construction of the stress-energy tensor will use in a fundamental way the notion of CLE on the Riemann sphere and on annular domains (doubly connected domains). This has not been developed yet, so we propose natural definitions from the known CLE construction on simply connected domains. We define a probability function for CLE on the Riemann sphere (definition \ref{defCplane}) and prove its global conformal invariance (theorem \ref{thglobinv}), under two conjectures about the CLE measure (assumptions \ref{asslimit}, about the measure on small loops, and \ref{assumpsym}, about a symmetry property). Then, we define the probability function for CLE on annular domains (definition \ref{nsc}). Under one additional conjecture about the CLE measure (assumption \ref{assborder}, about no loop touching one boundary of the domain of definition in the doubly connected case, a natural analogue of a property of the CLE measure on simply connected domains), we prove conformal invariance on these domains (theorem \ref{thcr3}), as well as three other theorems that will play a crucial r\^ole in the second part of the work (theorems \ref{thcr1}, \ref{thcr2} and \ref{corss}). We provide justifications for the three conjectures made. Although all results are obtained in view of the stress-energy tensor construction, we believe that the notions, theorems and techniques of the present paper are interesting in the more general CLE context.

{\bf Acknowledgments}

I am extremely grateful to J. Cardy for numerous discussions about this subject, and to W. Werner for sharing his knowledge and insight about CLE and for comments about the manuscript. I would also like to thank D. Bernard, P. Dorey, O. Hryniv and Y. Saint-Aubin for discussions, comments and interest, and I am grateful to D. Meier for reading through the  manuscript. I acknowledge the hospitality of the Centre de Recherche Math\'ematique de Montr\'eal (Qu\'ebec, Canada), where part of this work was done and which made many discussions possible (August 2008).

\sect{Conformal loop ensembles}

\subsection{Collective objects in the scaling limit and the need for CLE}

\begin{figure}
\bc
\includegraphics[width=7cm,height=7cm]{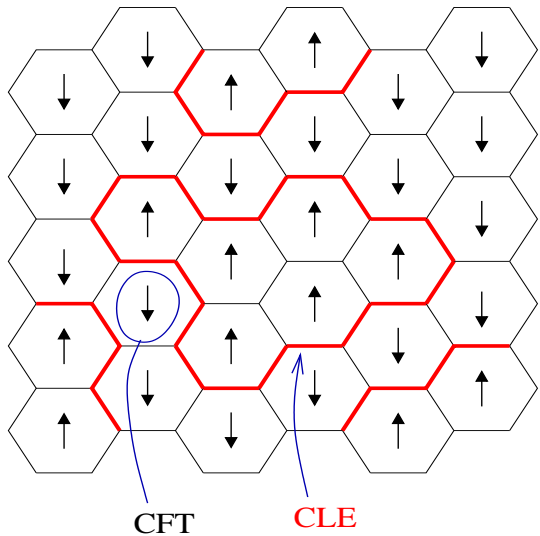}
\ec
\caption{Example of a few hexagonal lattice sites with Ising spins on the faces and the corresponding cluster boundaries. CFT describes most easily the fluctuations of the spins, and CLE that of the cluster boundaries.}
\label{figlattice}
\end{figure}

One of the most beautiful ideas that emerged in the context of critical systems is that of describing in the scaling limit, instead of the local statistical variables, the fluctuating {\em boundaries} of clusters of such variables, or other natural curves occuring from them, through measure theory on sets of loops and curves in the plane. For instance, in a model of magnetism where magnetic moments can point in only two directions, like the Ising model, one may form clusters of aligned moments (see figure \ref{figlattice}).
In this example, cluster boundaries in any given configuration are curves through which the moments flip. As alluded to in the introduction, these boundaries are the proper collective objects of CFT: large clusters are what represent collectivity best, and at criticality (or near to it), their boundaries are far enough apart to produce a set of curves in the scaling limit. The first successful measure theory for such curves was obtained by Schramm \cite{S00}. The idea of considering cluster boundaries as a way to provide a precise meaning of conformal invariance and universality was discussed earlier in \cite{LPSA94,LLSA00}, where the question was studied numerically. The power of the description in terms of random curves and loops comes from the fact that precise notions of conformal invariance and locality can be stated, leading to natural families of measures for these objects directly in the scaling limit: SLE \cite{S00} and CLE \cite{W05a,Sh06,ShW07a,ShW07b}.

It is interesting to note that in a sense, the description of the scaling limit through cluster boundaries is dual to that of CFT: the latter deals most easily with local statistical variables through the concept of local fields, whereas the former deals with extended objects. It is also interesting to keep in mind that there is another interpretation of the random loops and curves, through underlying models of quantum particles: they can be seen as trajectories of relativistic quantum particles (propagating in ``imaginary time'', since the signature is Euclidean), or perhaps as the dual surfaces perpendicular to these trajectories. This will be useful in understanding the meaning of the stress-energy tensor construction in the second part of this work.

\begin{figure}
\bc
\includegraphics[width=5cm,height=5cm]{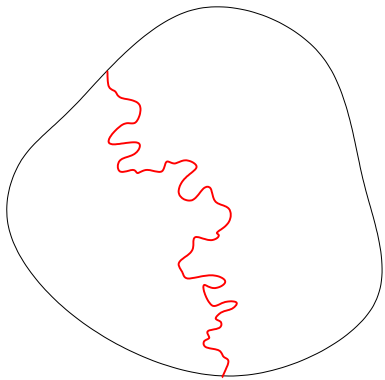}
\ec
\caption{Drawing representing an SLE curve on a domain.}
\label{figSLE}
\end{figure}

In SLE, one considers the situation where the system is on a domain (open set) of the Riemann sphere, and is set up such that there is a cluster boundary that starts and ends on the boundary of the domain. SLE describes the random fluctuations of this curve (see figure \ref{figSLE}).
For instance, in the previous example, the magnetic moment of the Ising model may be required to flip at exactly two points of the system boundary -- then, there is a unique cluster-boundary curve that starts and ends on the boundary of the domain. There is a continuous family of SLE measures, parametrised by $\kappa\in[0,8]$. Its few defining properties, conformal invariance and the domain Markov property, allowed the proof of the existence of the scaling limit in important models (first done in \cite{Smi01}).

Concerning constructive CFT, the relation between SLE and CFT has been developed to a large extent: works of the authors of \cite{BB06} reviewed there, and works \cite{FK04,F04,Du07,KS07}, considering the relation between CFT correlation functions, partition functions, and martingales of the stochastic process building the SLE curve; works \cite{FW02,FW03} considering the relation between the CFT Virasoro algebra on the boundary (the boundary stress-energy tensor) and local SLE variables\footnote{A {\em local} variable in SLE may be loosely understood as a variable that is unaffected by deformations of the SLE curve if the curve lies away from a given ``small'' region.}, the generalisation \cite{DRC06} to the bulk stress-energy tensor, and a related study of other bulk local fields in \cite{RC06}. From some of these works, it is known that SLE measures correspond to a continuum of central charges $c$ less than or equal to 1, with $c = (6-\kappa)(3\kappa-8)/(2\kappa)$, and that a large family of CFT correlation functions are associated with SLE martingales.

However, SLE is fundamentally limited from the viewpoint of constructive CFT. For instance, it cannot describe all correlation functions of local fields, in particular bulk fields, since one is restricted to the condition on the existence of the SLE curve itself. But more fundamentally, it does not provide a clear correspondence between local CFT fields and the underlying local statistical variables. This is because from the viewpoint of the construction via martingales, the CFT correlation functions are expectations of extremely non-local random variables of the SLE curve. In the context of SLE, a proof that certain statistical variables are described by CFT local fields would require a proof that their correlation functions become specific martingales in the scaling limit. Locality of fields and the multilinear structure of correlation functions become very unnatural concepts. There are local SLE variables: for instance, a Schramm event \cite{S01}, that the curve lies to the right of a given point. For such variables, we may consider averages of products, and reproduce multi-linear CFT correlation functions. However, the associated fields mostly fall outside of the rational CFT descriptions; for instance, the Schramm events are zero-dimensional in the CFT sense yet different from the identity. The only exceptions are the objects constructed in \cite{FW03,DRC06,RC06}, related to the stress-energy tensor and other particular rational fields, but are associated to particular values of $\kappa$. Finally, a complete understanding of more subtle CFT concepts, like the partition function (without boundary fields), probably cannot be obtained purely from SLE.

The reason for these difficulties is that SLE does not describe enough of the scaling limit, concentrating solely on one particular cluster boundary. We need to describe all cluster boundaries; these are all collective objects. In the Ising model, for instance, it is easy to define, on the lattice, the random local magnetic moment in terms of a local\footnote{More precisely, it is semi-local -- this will be briefly discussed in the second part of this work.} random variable of all cluster boundaries (see below), and correlation functions are indeed expectations of products of these. Also, the ``counting of states'' of QFT probably agrees with (an appropriate renormalisation of) the counting of configurations of all cluster boundaries, so that all cluster boundaries should be needed in order to study CFT partition functions.

The latter point is closely related to the construction of the stress-energy tensor. Physical intuition about statistical models suggests that the local distortions that generate space transformations should affect locally any cluster boundary. As the stress-energy tensor is a generator of conformal transformations, it is natural to expect that it can be seen as a random variable of all cluster boundaries, localised at a point. From the point of view of relativistic quantum particles, with cluster boundaries representing their Euclidean space-time trajectories, the stress-energy tensor, sometimes called the energy-momentum tensor, measures the energy and momentum of these particles at a space-time point, hence likewise should be sensitive to all trajectories, and localised at a point.

In \cite{DRC06} (results that generalised those of \cite{FW02,FW03} to the bulk stress-energy tensor), it was shown that for $\kappa=8/3$, the stress-energy tensor can be constructed in SLE as a local variable. This construction made strong use of the property of conformal restriction particular to $\kappa=8/3$ \cite{LSW03}, and lays support to the discussion above. The case $\kappa=8/3$ corresponds to the central charge $c=0$. There, physical intuition indicates that there is no energy in the vacuum in a quantum-model perspective, so that the energy is indeed supported only on the SLE curve itself (there is only one trajectory, no ``vacuum bubbles''). Also, the underlying statistical model leading to $\kappa=8/3$, the self-avoiding random walk, is indeed a ``local-interaction'' model of a random curve without the need for loops -- only the curve feels space transformations. Hence again, it is natural that the stress-energy tensor be supported on this curve. Clearly, then, a generalisation of the construction of \cite{DRC06} to non-zero central charges, hence taking into account the conformal anomaly, would require the inclusion of all cluster boundaries.

\begin{figure} 
\bc
\includegraphics[width=7cm,height=7cm]{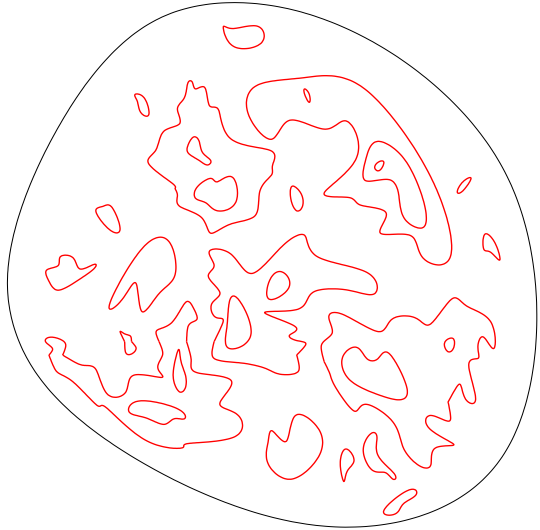}
\ec
\caption{Drawing representing a CLE loop configuration on a domain.}
\label{figCLE}
\end{figure}

The scaling limit of all cluster boundaries (and without the need for the SLE curve itself) is expected to give CLE. This provides a measure-theoretic description of all collective objects: unintersecting random loops in simply connected domains (see figure \ref{figCLE}) \cite{W05a,Sh06,ShW07a,ShW07b}.
There is a one-parameter family of CLE measures, CLE$_\kappa$ with $8/3<\kappa<8$ (the same $\kappa$ as in SLE, in a precise sense -- see below), hence expected to give central charges between 0 and 1. CLE is expected to describe the same universality classes as those of CFT for these central charges. There is a proof of convergence to CLE at $\kappa=6$ for the percolation model \cite{Smi06,CN06c,CN06a,CN06b}, and at $\kappa=16/3$ and $\kappa=3$ for the (dual versions of the) Ising model \cite{Smi06,Smi07,Smi08a,Smi08b,Smi08c}. In general the works \cite{W05a,Sh06,ShW07a,ShW07b} as well as the results of \cite{SSW06} and \cite{NW09} give precise descriptions of the random loops in all cases. Another construction is that of the Gaussian free field \cite{SS06}, for $\kappa=4$. The work \cite{KS07} also provides a discussion of the measure on all loops.  Concerning constructive CFT from CLE, the work \cite{Smi07} gives a candidate for the Ising holomorphic fermion, and a recent work proposed a way of obtaining the CFT local field corresponding to the local Ising magnetic moments from a CLE construction at $\kappa=16/3$ \cite{CN08}. However, it is still in general an open problem to identify random variables of the CLE loops with local CFT fields (many will fall outside of the rational description), and it is not clear if all rational CFT fields can be obtained in this way. In general, it is not clear what the concept of local fields means in CLE; we will attemps to provide some clarifications in the second part of this work.

\subsection{General aspects of CLE and relation to the scaling limit of the critical $O(n)$ models}

Conformal loop ensembles form a family of measures for random loops in the plane, with properties of conformal invariance. There are two very distinct regimes for conformal loop ensembles: a {\em dilute} regime where the measure is on random simple (without double or higher order points) loop configurations in a simply connected domain \cite{W05a,ShW07a,ShW07b}, and a {\em dense} regime where it is on random quasi-simple (with possible double points) loop configurations in the closure of a simply connected domain \cite{Sh06}. We will only consider the first regime. To be precise, a simple loop is a subset of the Riemann sphere $\C$ that is homeomorphic to the unit circle $S^1$, and a configuration in the dilute regime is a set of disjoint simple loops in $\C$ that is finite or countable. In the dilute regime of CLE, we look at configurations such that all loops lie in some simply connected domain. We will call this domain the {\em domain of definition} of the CLE. There is a measure for any simply connected domain of definition, and for any value of the parameter $\kappa\in(8/3,4]$.

In order to have a picture of why conformal loop ensembles may describe scaling limits, and why they may be related to CFT, it is useful to know that they are expected to represent, in the dilute regime, the scaling limit of the loops in the so-called $O(n)$ lattice loop models, for $0<n\leq 2$. An $O(n)$ loop model on a finite graph is a measure on configurations of disjoint loops on the graph. It is given by $x^s n^\ell$, where $x$ is a function of $n$ that depends on the graph, $s$ is the total length of all loops (the number of vertices that are part of a loop), and $\ell$ is the number of loops. The $n$-dependent parameter $x$ is chosen in such a way that the model is critical (something which, of course, can only be assessed by studying an appropriate infinite-graph limit, the so-called thermodynamic limit). It is known that, for instance, for the regular hexagonal lattice, we must choose $x = 1/\sqrt{2+\sqrt{2-n}}$ \cite{N82}. In the scaling limit, with two-dimensional graphs (which has a precise meaning in the limit of large graphs), these should then give rise to models of CFT.

The $O(n)$ lattice models include, for instance, the statistical Ising model at $n=1$, where the loops may be understood as representing boundaries of clusters of aligned spins (see figure \ref{figlattice}). These models are in fact expected to give rise to all minimal models of CFT in the scaling limit, by appropriately choosing $n$ in order to give a minimal-model central charge -- this is a countable family of models \cite{BPZ}. They also are expected to give rise to non-minimal models, where the usual module structure reduction does not apply. The full relation between the $O(n)$ models and models of CFT is expected to be recoverable using Coulomb-gas, or free-field, analysis (see for instance \cite{C00}). This provides candidates for primary fields through the so-called exponential fields, and possibly similar fields with additional topological properties (but we should note that the construction of \cite{CN08} is not of this type, since a case with $\kappa>4$ was studied). Essentially, the exponential fields are the scaling limit of random variables of the form $u^{\ell'}$ for some $u\in\C$ and where $\ell'$ is the number of loops that separate two points $p$ and $p'$ on the graph, or a point $p$ on the graph from the boundary vertices. They naturally generalise the spin field of the Ising model, which is just the case $u=-1$. They point to natural counterparts as random variables in CLE, but this is very non-trivial because there are infinitely many loops (one would need a renormalisation process). The Coulomb gas construction of CFT also gives a free-field form for the stress-energy tensor (see, for instance, \cite{DFMS97}), which may be seen as suggesting a representation in terms of random variables on the loops in CLE. However, it is also an extremely non-trivial matter to make this precise and provide a proof.

As we mentioned in the introduction, in this work, we will show that there is a way of constructing a local CLE ``object'' that describes the stress-energy tensor, for any CLE$_\kappa$ with $8/3<\kappa\leq4$. One of the conclusions from this construction, presented in the discussion section of the second part, will be a precise statement about the free-field form of the stress-energy tensor in CLE, derived entirely in the CLE context. An analysis of this construction will also provide insight into the structure of local fields and the meaning of the partition function in the CLE context.

\subsection{The axioms of the CLE measure, and their interpretation}\label{subsectaxioms}

We now introduce the more technical aspects of CLE, and refer the reader to appendix \ref{notations} for some notations and conventions used in this section and throughout this paper.

First, we make more precise the configuration space by stating the two ``finiteness'' properties satisfied by the loops that are found in any configuration. The first property was discussed above, but is expressed here in a more explicit fashion. The second is an additional property, telling us how the loops in any configuration can be counted.

In order to state these properties, we introduce the concept of radius of a set. For us, the {\em radius} of a set in some simply connected domain $C$ is the minimal radius of a closed disk in $\uD$ that covers the set, when this set is mapped from $C$ to the unit disk $\uD$ (hence the radius is always between 0 and 1). In order to make it unique, we just choose a conformal transformation $C\to\uD$ for any given $C$ (there is always such a conformal transformation by Riemann's mapping theorem). See appendix \ref{notations}. We will also use the word {\em extent} to represent the corresponding diameter, i.e.\ twice the radius, and the phrase {\em distance between two points} to represent the extent of the two-point set. The radius is certainly a $C$-dependent quantity: it depends on the simply connected domain where the set lies. This domain is taken, unless stated otherwise, as the domain of definition of the CLE under consideration. Also, the radius is {\em not} a conformally invariant quantity.

The two properties defining the set of configurations are as follows.
\begin{itemize}
\item {\bf Finiteness I.} In any configuration, there is a finite number or a countable infinity of loops, and all loops are simple, do not have points in common with each other, and do not have points in common with the boundary of the domain of definition. That is, 1) if the extent of any part of a loop between two points on the loop is non-zero, then the two points are also a non-zero distance apart; 2) the distance between any two loops is non-zero; and 3) the distance between any loop and the boundary of the domain is also non-zero.
\item {\bf Finiteness II.} In any configuration, the number of loops of radius at least $d$ is finite for any given $0<d<1$.
\end{itemize}
An immediate consequence of the finiteness II property is that if there are infinitely many loops in some configuration, then the loops  can be counted by visiting them in order of decreasing radius -- the set of loops is open at the ``small-loop end'' only. Hence this precludes ``accumulations'' of loops: it is not possible to scan through all loops of radius greater than a fixed number $d$ and get a smaller and smaller distance to another fixed loop. In the set of loops of radius at least $d$, the set of distances between loops has a minimum greater than 0, for any $0<d<1$. However, as we look at decreasing $d\to0$, this minimum may well (and in fact does) decrease to 0.

Some direct implications are for instance as follows. In any configuration, for $B$ and $B'$ two simply connected domains with $B\subset B'\subseteq C$, the number of loops that surround $B$ and are included inside $B'$ is finite. This is because this number of loops is less than the number of loops of radius at least that of $B$. The latter number is finite since any domain has a non-zero radius. Also, the number of loops that intersect two domains $B\subset C$ and $B'\subset C$ whose closures do not intersect is finite. This is because this number of loops is less than the number of loops of extent at least the distance between $B$ and $B'$. That distance is non-zero since any two domains whose closures do not intersect are a non-zero distance apart.

We now define the probability space of CLE. For any simply connected domain $C$, we consider a probability space $(\conf_C,\sgm_C,\mu_C)$, where $\conf_C$ is the set of configurations on $C$, $\sgm_C$ is the associated $\sigma$-algebra, and $\mu_C$ is a CLE measure on $\sgm_C$.

We recall \cite{H50} that a $\sigma$-algebra is a set of events closed under negation and countable unions and containing the {\em trivial event} $\conf_C$. An event is defined as a subset of the set of configurations (that is, here, an element of the set $\evs(\conf_C)$ of subsets of $\conf_C$). A particular event can be specified by expressing a set of properties that are satisfied by all configurations that belong to the event but none of the configurations that do not belong to it. We will use the phrase {\em the event that} when specifying an event in this way. We will also use the phrase {\em evaluation of the event in a configuration} for the procedure of checking if this configuration is element of the event.

As for the $\sigma$-algebra used in CLE, precise definitions of $\sgm_C$ can be found in, for instance, \cite{Sh06,ShW07a}. Consider for instance $C=\uD$ (the unit disk). First put a metric structure on the space of simple loops in $C$ through the Hausdorff distance\footnote{The Hausdorff distance between two subsets $A$ and $B$ of a metric space with distance function $d$ is $D(A,B) = {\rm max} ({\rm sup}_{x_1\in A} {\rm inf}_{x_2\in B} d(x_1,x_2),\,{\rm sup}_{x_1\in B} {\rm inf}_{x_2\in A} d(x_1,x_2))$.} between two loops induced by the Euclidean distance on $\R^2$. Then, put a metric structure on the space of configurations through the Hausdorff distance between two configurations induced by the metric on the space of loops. In this metric space of configurations, one considers the $\sigma$-algebra of Borel subsets (essentially, generated by ``higher-dimensional'' intervals). It will be convenient sometimes to recall this metric space on configurations, but mostly, we will just consider events as conditions on ``big enough'' loops; for instance, the events that exactly $n$ loops are present that intersect simultaneously $m$ sets whose closures are pairwise disjoint, for $n=0,1,2,3,\ldots,\,m=2,3,4,\ldots$, as well as similar events with additional topological conditions, for instance that the loops separate two points.

\begin{figure}
\bc
\includegraphics[width=10cm,height=7cm]{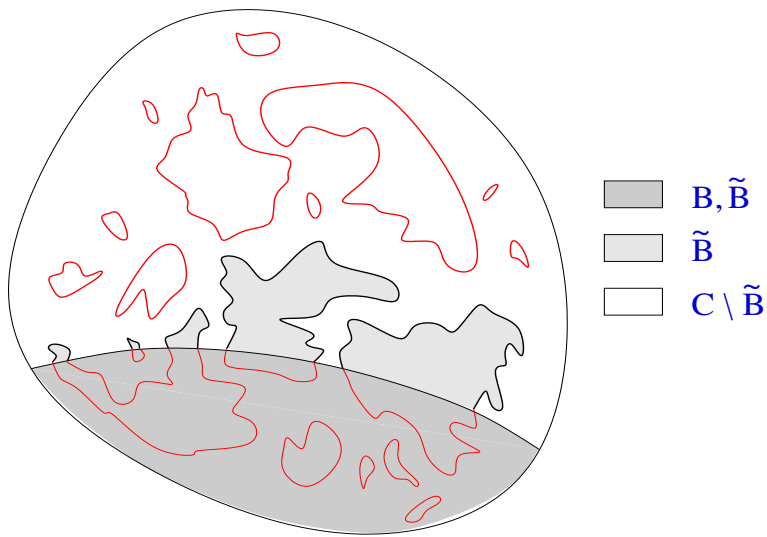}
\ec
\caption{The various domains involved in conformal restriction, from the outer loops of the configuration depicted in figure \ref{figCLE}.}
\label{figCLErest}
\end{figure}

A family of CLE measures $\mu_C$ in the dilute regime, parametrised by simply connected domains $C$, is defined by the following properties \cite{W05a,ShW07a}:
\begin{itemize}
\item {\bf Conformal invariance.} For any conformal transformation $g:C\to C'$, we have $\mu_C = \mu_{C'}\circ g$ (where $g$ is applied individually to all configurations of the event, and there individually to all loops of these configurations).
\item {\bf Nesting.} Consider an outer loop $\gamma$ of a configuration (a loop that is not inside any other loop) and the associated domain $C_{\gamma}$ delimited by $\gamma$ and lying in $C$ (that is, $C_\gamma$ is the interior of the loop $\gamma$ in $C$). The measure $\mu_C$ conditioned on all outer loops $\gamma$ (this is a countable set), as a measure on $\conf_{\cup_\gamma C_\gamma}$, is a product of CLE measures on each individual interior domain, $\otimes_\gamma \mu_{C_\gamma}$. See below for the meaning of {\em conditioning}.
\item {\bf Conformal restriction.} Given a domain $B\subset C$ such that $C\setminus B$ is simply connected, consider $\t{B}$, the closure of the set of points of $B$ and points that lie inside loops that intersect $B$. Consider also the connected components $C_j$ of $C\setminus \t{B}$ ($j$ is in a countable set). Then the measure $\mu_C$ conditioned on all loops that intersect $B$ or lie inside $\t{B}$, as a measure on $\conf_{\cup_j C_j}$, is a product of CLE measures on each individual components, $\otimes_j \mu_{C_j}$. We will call this the {\em restriction based on $C \setminus B$} and call $C\setminus \t{B}$ the {\em actual domain of restriction}. See figure \ref{figCLErest}.
\end{itemize}

These axioms make reference to the process of {\em conditioning} on random objects (e.g.\ outer loops) in order to obtain a measure on other random objects (e.g.\ the loops inside outer loops). This process involves, essentially, looking at every configuration weighted according to the initial measure (e.g.\ $\mu_C$), and in each configuration, re-randomising the objects that are not conditioned on. The statement that upon conditioning a certain measure is obtained on the latter objects (e.g.\ $\otimes_\gamma \mu_{C_\gamma}$), says that if this re-randomisation is made according that measure, then the resulting measure on all configurations is again the initial measure. What is important is that the conditioning should be on objects that are identifiable in any configuration. For instance, one cannot condition on ``a loop having such and such shape.'' Above, the outer loops in the nesting axiom are well identifiable, as well as the loops involved in the conformal restriction axiom.

It is possible to re-state the nesting and conformal restriction axioms by involving, instead of all outer loops or all connected components of an actual domain of restriction, only one of them, chosen in an appropriate way. For the nesting property, we first repeat the idea of nesting but for loops inside outer loops, etc., so that we may consider the measure for the interior of any loop, not just outer loops (as long as there is a precise procedure to choose it in any configuration). Then, taking the measure conditioned on some chosen loop and its exterior, we obtain a CLE measure in its interior, as long as the choice is made without the information of the interior loops. A way of guaranteeing this is by discovering loops in a sequence where no loop surrounds an earlier loop, stopping at the chosen one. Something similar holds for conformal restriction, where the chosen component has to be such that it never contains any of the loops sequentially discovered in order to make the choice. In the next subsection we will provide a re-statement of the axioms above, making more explicit the conditioning on random objects involved and the way in which choices may be made.

The axioms above have very natural interpretations. The property of conformal invariance is the main statement of criticality of the lattice model. From the viewpoint of the lattice $O(n)$ model, it is essentially the only one that needs a non-trivial proof. The requirement that conformal invariance holds for any simply connected domain implies a very special structure of the underlying statistical model: it involves not only global scaling transformations, but also local ones, as the domain can have essentially any shape. Hence, this requirement encodes both the lack of a scale at criticality, and a certain local aspect of hypothetical underlying statistical models with localised fluctuating variables. However, we are looking at models of loops, which are extended objects. The other two properties can be seen as completing the expression of locality for such extended objects. Interpreted on the lattice, they are immediate consequences of the measure in the $O(n)$ model, at or away from criticality. That is, they follow from 1) the product form of the measure in the lattice $O(n)$ model, $x^s n^\ell = \prod_\gamma \omega(\gamma)$, $\omega(\gamma) = nx^{|\gamma|}$ where $|\gamma|$ is the length of the loop $\gamma$ in the configuration, and 2) the constraint of having disjoint loops in the configuration space. Indeed, the conditioning on loops, in this measure, simply divides out the factors corresponding to these loops, and the rest is a product of measures all of which have the same product form, but with the restriction that loops lie in smaller simply connected domains. This is just the product of $O(n)$ measures on smaller simply connected domains, as in nesting or conformal restriction. In QFT terms, the product form implies a Hamiltonian that is a sum over individual loops, and the constraint of disjoint loops gives ``ultra-local'' repulsion terms amongst them. Of course, the loops are themselves extended objects, but this form of the Hamiltonian is certainly the most local we may impose for such objects.

Let us now state the nesting and conformal restriction axioms in more intuitive terms. Nesting is simply the statement that the interior side of a loop is itself like the boundary of a CLE domain of definition. On the other hand, conformal restriction is an ``attempt'' at two statements: 1) that the exterior side of a loop is also like the boundary of a CLE domain of definition, and 2) that if all loops are restricted not to intersect $\p B$ (where $B$ is the domain in the statement of conformal restriction), then $C\setminus \cl{B}$ ($\cl{B}$ is the closure of $B$) is a new CLE domain of definition. The first statement is obviously natural, but the second also is: if no loop intersects $\p B$, then $C\setminus \cl{B}$ is ``separated'' from $B$, so its configurations are independent from the loops in $B$ and it forms a new CLE domain of definition. Conformal restriction as stated above would be a consequence of these two statements put together. In the lattice case, both statements are direct consequences of disjointness of the loops and of the product form of the measure in the lattice $O(n)$ model. However, none of them can be imposed on CLE measures. The first one cannot be imposed, because we would need to extend the family of CLE measures to multiply connected domains of definition. The second cannot for a more subtle reason: it is impossible to restrict the measure to no loop interesecting $\p B$, since almost surely, almost all points are surrounded by a loop  -- see below for properties of CLE (the phrase {\em almost surely} applied to an event indicates that the negation of the event has measure zero). Only the weaker statement of conformal restriction above may be imposed.

The usefulness of these CLE axioms relies in great part on a strong uniqueness theorem \cite{ShW07a}. By the property of conformal invariance, we may restrict our attention to one given domain of definition, say the unit disk $\uD$, and by the property of nesting, we may restrict our attention to the outer loops in any configuration on $\uD$. Then, conformal invariance for transformations preserving $\uD$ and conformal restriction give constraints on $\mu_\uD$ as a measure for these outer loops. Note that there are ``few'' conformal transformations preserving $\uD$: they form the $SU(1,1)$ group. Hence, here conformal invariance is not such a strong constraint by itself. However, conformal restriction is very strong. It is shown in \cite{ShW07a} that there is at most a one-parameter family of measures $\mu_\uD$ on $\conf_\uD$ that satisfies these constraints. Including all nested loops again, it has the property that in any configuration, there is almost surely a countable infinity of loops. The loops controlled by these CLE measures ``look like,'' locally, parts of SLE$_\kappa$ curves for some $\kappa$. The family can be parametrised by this $\kappa$, and it turns out that all possibilities are exhausted with $8/3<\kappa\leq4$ (where the SLE curve is simple), as mentioned above. These CLE measures are constructed in \cite{ShW07b}.

In \cite{ShW07a}, the finiteness II property above is expressed as an additional axiom of the CLE measure instead of being taken as a property of the configuration space (the axiom is that finiteness II is satisfied almost surely). In fact \cite{ShW07a}, finiteness II could probably be omitted altogether, and proved from the three axioms above. However, this is apparently a rather hard proof, and in any case, finiteness II should be a natural outcome of the ``usual'' proof, when it is available, of the existence of the continuum conformally invariant limit of a lattice measure.

Recall the finiteness I property: that loops in configurations of $\conf_C$ are simple and disjoint from each other and from the boundary of the domain of definition. Consider a completion of $\conf_C$ under finiteness I: configurations where loops do not cross themselves or each other, but are allowed to have double points and points in common with each other and with the boundary. Then, the same family of CLE measures is found from the three axioms above, with almost surely simple, disjoint loops, disjoint from the boundary. But as we mentioned, there is another CLE formulation for the completion of $\conf_C$ \cite{Sh06}. It gives $4<\kappa<8$ (the dense regime), where the completion of finiteness I holds almost surely. For $\kappa=4$, another formulation is that of the Gaussian field \cite{SS06}. As $\kappa$ approaches $8/3$, the ``density'' of loops decreases. At $\kappa=8/3$ there are no loops remaining, but this case could also be taken as a measure on configurations of only one self-avoiding loop \cite{W05b}. For $\kappa<8/3$, there is no theory of loop ensembles. In this paper, we will only consider the cases $8/3<\kappa\leq4$.

\subsection{Probability function and re-randomisation procedures} \label{ssectrandom}

The measure discussed above induces a probability function for events in the $\sigma$-algebra. This is not entirely trivial, because the CLE measure is an infinite measure, due mainly to conformal invariance: conformal transformations preserving a simply connected domain $C$ form a non-compact group. A way this can be thought of is by considering the measure induced by $\mu_C$ on all loops that whose radius is greater than a certain minimal radius. The radius is {\em not} a conformally invariant quantity; under conformal transformations, small loops can become big loops, and {\em vice versa}. Hence, the induced measure on big enough loops is finite, because conformal invariance is broken. Then, we may make this induced measure a probability measure. Yet, for any given event of $\sgm_C$, the probability it induces is independent of the minimal radius chosen if this radius is small enough, because small loops do not affect the evaluation of the event. For instance, the evaluation of the event that at least one loop intersects two domains a finite distance apart, is not affected by excluding the very small loops from a CLE configuration. Hence the limit of the probability when the minimal radius goes to zero exists. This is the CLE probability for this event.

The CLE probability function on the simply connected domain $C$ will be denoted $P(\tou)_C$, for $\tou \in\sgm_C$. Conjunction of events will be separated by a comma, for instance: $P(\tou \cap \t\tou)_C = P(\tou,\t\tou)_C$. The probability of an event $\tou$ restricted to another event $\tou'$ will be denoted $P(\tou|\tou')_C$. In fact, instead of considering a different set of events $\sgm_C$ for every simply connected domain $C$, it will be convenient to consider events as general elements of $\evs(\conf_\C)$, subsets of the set of configurations on the Riemann sphere $\C$. For $\tou\in\evs(\conf_\C)$, we consider $\tou_C\equiv \tou\cap \conf_C$, the restriction of $\tou$ on $\conf_C$. If $\tou_C\in\sgm_C$, then we simply define $P(\tou)_C \equiv P(\tou_C)_C$. The fact that we consider events as subsets of $\conf_\C$ means that we cannot obtain, by restriction on $\conf_C$, events in $\sgm_C$ that make ``explicit reference'' to the boundary of the domain. For instance, the event that a loop surrounds a certain domain is excluded: the concept of surrounding is $C$-dependent, and in general ill-defined on $\C$. This turns out to be crucial when we define probability functions on $\C$. Finally, the negation of an event $\tou$ will be denoted by $\non\tou = \conf_\C\setminus \tou$. 

A way of interpreting the probability function induced from the measure $\mu_C$ is to consider an infinite sequence of configurations, a {\em time-sequence}, imagining that we draw configurations at random and evaluate all averages and probabilities in $\sgm_C$ from summing over an infinity of these draws. Since we have a probability function on the $\sigma$-algebra, this can always be done by the large-number theorem. For a sequence $\sam$, we will denote the probability function by $P_\sam$ (see (\ref{Psam})). Not any sequence can be considered to be such a time-sequence: every draw is independent and equivalent, and there is ergodicity. For instance, a property that is satisfied by all configurations of the time-sequence is a property that is almost sure for the measure $\mu_C$. If for instance a random integer $n\ge0$ evaluated from the random configurations is almost surely finite, then we must have $\lim_{N\to\infty} \sum_{n=0}^N P_\sam(n) =1$ (in a slight abuse of notation for $P_\sam$). Of course, not all infinite sequences whose elements have finite $n$ satisfy this condition (take for instance a sequence $\sam$ that gives the values $n=1,\,2,\,3,\,\ldots$: there, we have $P_{\sam}(n)=0$ for any finite $n$), but it is satisfied by infinite sequences of randomly generated independent configurations whose elements all have finite $n$.

We find the viewpoint of time-sequences useful especially in interpreting and using the defining properties of CLE. Let us fix a family of such infinite sequences,
\[
    \sam_C=(x_1^C,x_2^C,\ldots),\quad x_j^C \in \conf_C,
\]
parametrised by simply connected domains $C$, reproducing the CLE measures:
\beq
    P(\tou)_C = P_{\sam_C}(\tou)
\eeq
for $\tou_C\in\sgm_C$.

The three main properties of the CLE measure can be translated into three properties of the family of functions $P_{\sam_C}$. Consider  $\tou_C\in\sgm_C$.
\begin{itemize}
\item {\bf Conformal invariance.} The conformal invariance property implies that for any conformal transformation $g:C\to C'$, we have
\beq\label{confinvprop}
    P_{\sam_{g(C)}}(g(\tou_C)) = P_{\sam_C}(\tou_C) = P_{\sam_C}(\tou).
\eeq
Note that we need to restrict $\tou$ to $\tou_C$ before taking its conformal image under $g$, since $g$ is conformal, a priori, only on $C$.

\item {\bf Nesting.} The nesting property implies that any configuration $x_j^C$ in the sequence can be replaced with a configuration where the loops inside a given loop are re-randomised, as long as this loop is ``chosen'' independently from the loops inside it. First, consider a given configuration $x$ and a loop $\gamma\subset x$. Let us construct the sequence $\sam_C'(\gamma,x)$ of configurations in $\conf_C$ by adjoining to every element of $\sam_{C_\gamma}$ the loop $\gamma$ and the set of loops in the exterior of $\gamma$, that is, $\{\gamma\} \cup \exter_{C_\gamma}(x)$ (see the definitions in appendix \ref{notations}). Second, consider a ``choice'' map $\Phi$, mapping configurations in a subset of $\conf_C$ to loops in $C$. It has the properties that for $x\in\conf_C$, if $\Phi(x)=\gamma$ is defined, then 1) $\gamma\in x$ (that is, we choose a loop in $x$), and 2) $\Phi(x')=\gamma$ for any $x'\in\conf_C$ such that $\gamma\in x'$ and $\exter_{C_\gamma}(x) = \exter_{C_\gamma}(x')$ (that is, the choice does not depend on what is in the interior of the chosen loop).

Let us consider the subset $M$ of all integers $m$ for which $\Phi(x_m^C)=\gamma_m$ is defined. Then, we have
\beq\label{nestprop}
    P_{\sam_C}(\tou) = \lim_{N\to\infty} N^{-1}\sum_{m=1}^N
    \lt\{\ba{ll} \pres(x_m^C,\tou) & (m\not\in M) \\
    P_{\sam_C'(\gamma_m,x_m^C)}(\tou) & (m\in M) \ea\rt.
\eeq
where $\pres$ is the characteristic function (\ref{pres}). In words, for every configuration where we chose a loop $\gamma_m$, we re-randomise the loops inside it as if they were independent CLE configurations on domains delimited by the loop $\gamma_m$, keeping the exterior loops intact. The re-randomisation procedure has the effect of allowing us to replace the characteristic function by a probability function.

\item {\bf Conformal restriction.} The conformal restriction property says something similar, except
that it is the simply connected components of the actual
domain of restriction $C\setminus \t{B}$ where we may re-randomise.
Consider $y(x)$ the set of simply connected components (that were denoted $C_j$ in subsection \ref{subsectaxioms}) of the actual domain of restriction obtained from $x\in\conf_C$. For a given configuration $x$, consider all loops $\Gamma\subset x$ not intersecting some chosen component $A\in y(x)$ (some of these loops form part of the boundary of $A$). Let us construct the sequence
$\sam_C''(A,x)$ by adjoining to every element of
$\sam_{A}$ the set of loops $\Gamma$. Consider also, as for the nesting property, a ``choice'' map $\Psi$, mapping configurations in a subset of $\conf_C$ to simply connected domains in $C$. It has the properties that for $x\in\conf_C$, if $\Psi(x)=A$ is defined, then 1) $A\in y(x)$, and 2) $\Psi(x')=A$ for any $x'\in\conf_C$ such that $A\in y(x')$ and $\exter_{A}(x) = \exter_{A}(x')$.

Now let us consider the subset $M$ of all integers $m$ for which $\Psi(x_m^C)=A_m$ is defined. Then, we have
\beq\label{confrestprop}
    P_{\sam_C}(\tou) = \lim_{N\to\infty} N^{-1}\sum_{m=1}^N
    \lt\{\ba{ll} \pres(x_m^C,\tou) & (m\not\in M) \\
    P_{\sam_C''(A_m,x_m^C)}(\tou) & (m\in M). \ea\rt.
\eeq
\end{itemize}

We will in fact also need the {\em restricted} versions of the re-randomisation procedures above. That is, we may evaluate $P_{\sam_C}(\tou|\tou')$ by taking the ratio $P_{\sam_C}(\tou,\tou')/P_{\sam_C}(\tou')$, but we may also evaluate it by restricting the summation variable $m$ above to the subset of the positive integers such that $x_m^C$ satisfies the conditions of $\tou'$. There, we may replace, instead, $\pres(x_m^C,\tou)$ by the restricted re-randomised version, $P_{\sam_C'(\gamma_m,x_m^C)}(\tou|\tou')$ or $P_{\sam_C''(A_m,x_m^C)}(\tou|\tou')$.

The properties of the choice functions above essentially tell us that the choice must be made independently from the loops in the domains where re-randomisation is performed. For instance, in the case of the nesting axiom, if a loop $\gamma$ was chosen in a configuration, then any other configuration where only the interior of $\gamma$ differs is also a configuration where the loop $\gamma$ is chosen. The choice functions should also have certain properties of continuity, but we will not go into these details here.

In one construction below, we will need a slightly more general re-randomisation procedure in the case of the nesting axiom. We will perform a re-randomisation whereby the loop $\gamma$, delimiting the domain of re-randomisation, is chosen in such a way that the configuration of loops inside it satisfies the conditions of a certain event $\toua$. Then, the resulting measure for the configurations of loops inside this loop is a CLE measure restricted on $\toua$. For simplicity, we will not express this procedure in its greatest generality, but we will only prove it in the case that we are interested in below, in subsection \ref{subsectsupport}.

Note that it is also possible to choose a set of loops in the exterior of each other, or a set of simply connected components of an actual domain of restriction, instead of just one loop or just one component, where we perform the re-randomisation; but we will not need this here.

It is in conjunction with the concept of support that the above re-statement of the CLE axioms will be most useful.

\section{Basic properties of CLE, continuity and support}

\subsection{Some basic properties}

We state here one basic proposition and two consequences. The proposition says that almost surely, almost every point is surrounded by at least one loop \cite{W05a,ShW07a}.
\begin{propo} \label{propoeverypoint} In any configuration of $\sam_C$, the Lebesgue measure on the set of points that are not surrounded by a loop is zero.
\end{propo}
By conformal invariance, this implies that it is possible, for any given point in $C$, to choose a time-sequence of configurations $\sam_C$ such that this point is surrounded by at least one loop in all configurations of $\sam_C$. Combined with the nesting property, this also implies that in any configuration of $\sam_C$ there is a countable infinity of loops around most points.

Another implication is that the probability that at least one loop surrounds a domain goes to 1 as the domain is made smaller and smaller. It is simplest and sufficient to express the latter property for CLE on $\uD$:
\begin{corol}\label{corolsw}
With $B_\delta\subset C$ a family of domains parametrised by their radii $\delta>0$, the probability, on $\uD$, that at least one loop surrounds $B_\delta$ has the limit $1$ as $\delta\to0$.
\end{corol}
\proof Thanks to proposition \ref{propoeverypoint}, there is almost surely one loop that surrounds at least one point in $B_\delta$. Let us consider the random variable $s$: the radius of the largest disk, whose center is in $B_\delta$, that does not intersect the largest of such loops; and $d\rho(s)$ the measure for this random variable (on Borel subsets of the interval $[0,1]$). Since this variable is almost surely non-zero (because loops are simple), we have $\lim_{t\to0} \int_{t}^1 d\rho(s) = 1$. But the probability $P_\delta$ that at least one loop surrounds $B_\delta$, is greater than or equal to the probability that $s>\delta$, that is, $P_\delta \geq \int_{\delta}^1d\rho(s)$. Hence, $\lim_{\delta\to0}P_\delta \geq 1$, which shows the corollary since also $P_\delta\leq 1$ for all $\delta$.
\eproof

Finally, another consequence of proposition \ref{propoeverypoint} and of the finiteness II property is a slight, but useful, strengthening of corollary \ref{corolsw}:
\begin{corol}\label{corolsurr}
With $B_\delta\subset C$ a family of domains parametrised by their radii $\delta>0$, the probability, on $\uD$, that at least one loop of radius less than $d$ surrounds $B_\delta$ has the limit $1$ as $\delta\to0$, for any $d>0$.
\end{corol}
\proof Thanks to proposition \ref{propoeverypoint} and to the nesting property, we can find a point of $B_\delta$ that is almost surely surrounded by infinitely many loops, for any $\delta>0$. But since there are only a finite number of loops of finite radius in any configuration of $\sam_\uD$ (finiteness II property), there are infinitely many loops of radius smaller than $d>0$. The rest of the argument goes along the lines of the argument for corollary \ref{corolsw}. Let us consider the random variable $s$: the radius of the largest disk, whose center is in $B_\delta$, that does not intersect the largest of such loops; and $d\rho(s)$ the measure for this random variable. Since this variable is almost surely non-zero, we have $\lim_{t\to0} \int_{t}^1 d\rho(s) = 1$. But the probability $P_\delta$ that at least one loop surrounds $B_\delta$, is greater than or equal to the probability that $s>\delta$, that is, $P_\delta \geq \int_{\delta}^1d\rho(s)$. Hence, $\lim_{\delta\to0}P_\delta \geq 1$, which shows the corollary since also $P_\delta\leq 1$ for all $\delta$.
\eproof

\subsection{Continuity and Lipschitz continuity}\label{subsectcont}

It will be crucial in the next section to have certain properties of continuity under smooth ``deformations'' of events, and later on in this work, in fact, to have differentiability. In the present paper, we will not consider differentiability; it will be discussed in the second part of this work. In this subsection, we define general notions of continuity and Lipschitz continuity, and prove some general theorems related to these concepts. In particular, we show how to guarantee continuity of a whole $\sigma$-algebra of events from properties of generating events.

Our main notion of continuity is that upon deformations of the domain of definition that tend to the identity, we recover the probability on the domain of definition:
\begin{defi}\label{defcont}
An event $\tou$ is continuous at the simply connected domain $C$ if
\beq
	\lim_{n\to\infty} (P(\tou)_{g_n(C)} - P(\tou)_C) = 0,
\eeq
for any sequence of transformations $g_n$ conformal on $C$ with $\lim_{n\to\infty} g_n = \id$ on $\cl{C}$.
\end{defi}
In this definition and other definitions below, it is implicit that the restriction of $\tou$ to $g_n(C)$ must be an event in $\sgm_{g_n(C)}$, and similarly for the restriction to $C$. Also, a sequence of transformations $g_n$ conformal on $C$ with $\infty\not\in \cl{C}$ is said to {\em tend to the identity} $\id$ on $\cl{C}$ if:
\[
	\forall \delta>0:\; \exists N \;|\; \forall n>N:\; |g_n(z)-z|<\delta\;\forall z\in\b{C}.
\]
If $\infty\in\b{C}$, then the sequence $g_n$ is said to tend to the identity on $\cl{C}$ if the sequence $g_n \circ f^{-1}$ tends to the identity on $f(\cl{C})$ for any conformal transformation $f:C\to B$ with $\infty\not\in \cl{B}$.

More generally, it will be convenient to a have a concept of continuity for any function of simply connected domains:
\begin{defi}\label{defcontgen}
A function $F$ on a space of simply connected domains is continuous at $C$ if
\beq
	\lim_{n\to\infty} (F(g_n(C)) - F(C)) = 0,
\eeq
for any sequence of transformations $g_n$ conformal on $C$ with $\lim_{n\to\infty} g_n = \id$ on $\cl{C}$.
\end{defi}
Note that the sequence $g_n(C)$ of simply connected domains indeed converges to $C$ under the natural Hausdorff metric on the space of domains (as long as $\overline{C}$ excludes $\infty$).

In order to prove continuity for given events, there is a useful intermediate step, which amounts to proving what we call strong continuity:
\begin{defi}\label{defscont}
An event $\tou$ is strongly continuous at the simply connected domain $C$ if
\beq
	\lim_{n\to\infty} P(\tou,\non g^{-1}_n\tou_{g_n(C)})_C =0
	\ \mbox{ and }\ \lim_{n\to\infty} P(g^{-1}_n \tou_{g_n(C)},\non \tou)_C=0
\eeq
for any sequence of transformations $g_n$ conformal on $C$ with $\lim_{n\to\infty} g_n = \id$ on $\cl{C}$.
\end{defi}

The relation between strong continuity and continuity is the following simple theorem:
\begin{theorem}\label{theocont}
An event that is strongly continuous at $C$ is continous at $C$.
\end{theorem}
\proof We simply write
\beqa
	\lefteqn{\lim_{n\to\infty} (P(\tou)_{g_n(C)} - P(\tou)_C) =} \n &=& \lim_{n\to\infty} (P(\tou_n)_{C} - P(\tou)_C) \n
		&=&  \lim_{n\to\infty} (P(\tou_n,\tou)_{C} + P(\tou_n,\non \tou)_{C} -
		P(\tou,\tou_n)_C - P(\tou,\non \tou_n)_C) \n
		&=&  \lim_{n\to\infty} (P(\tou_n,\non \tou)_{C} -
		P(\tou,\non \tou_n)_C) \n
		&=&0
\eeqa
where $\tou_n \equiv g_n^{-1}\tou_{g_n(C)}$, and the last step is by strong continuity.
\eproof

Given that we have continuity for a family of events, it is useful to know if continuity holds also for events formed out of these by unions, intersections, etc., and more generally for the whole $\sigma$-algebra generated from it. Continuity of unions, for instance, does not follow from continuity of the individual events, except if these events are disjoint. However, we can prove that strong continuity of a family of events implies strong continuity, hence continuity, for the whole $\sigma$-algebra, which will play an important r\^ole below.
\begin{theorem}\label{theogener}
If all events in the set $\evs$ are strongly continuous at $C$, then all events in the $\sigma$-algebra generated from $\evs$ are also strongly continuous at $C$.
\end{theorem}
\proof In order to generate a $\sigma$-algebra, we adjoin the trivial event $\conf_\C$ (which is obviously strongly continuous) to $\evs$ if it is not already there, and consider all events generated under negation and countable unions. First, it is clear that under negation strong continuity is preserved: the definition is unchanged. Let us consider a sequence of strongly continuous events $\tou^{(i)},\,i\in\N$ and consider the countable union $\tou \equiv \cup_i \tou^{(i)}$. Let us write $\tou_n^{(i)} \equiv g_n^{-1}\tou_{g_n(C)}^{(i)}$ and likewise for $\tou_n$. We have
\beqa
	\tou \cap \non\tou_n &=& (\cup_i \tou^{(i)}) \cap \non(\cup_j \tou_n^{(j)}) = (\cup_i \tou^{(i)}) \cap \cap_j\non\tou_n^{(j)} \n
		&=& \cup_i (\tou^{(i)} \cap \cap_j \non\tou_n^{(j)}) \subseteq \cup_i(\tou^{(i)} \cap \non\tou_n^{(i)})
		= \cup_i \toua_n^{(i)} \no
\eeqa
where in the last step we defined $\toua_n^{(i)}=\tou^{(i)} \cap \non\tou_n^{(i)}$. Let us also define $\toua_n = \cup_i\toua_n^{(i)}$. Likewise, we have
\beqa
	\non\tou \cap \tou_n &=& \cap_i \non\tou^{(i)} \cap (\cup_j \tou_n^{(j)})
		= \cup_j (\cap_i \non\tou^{(i)} \cap \tou_n^{(j)}) \n
		&\subseteq& \cup_j(\non\tou^{(j)} \cap \tou_n^{(j)})
		= \cup_j \t\toua_n^{(j)}, \no
\eeqa
with $\t\toua_n^{(i)} = \non\tou^{(i)} \cap \tou_n^{(i)}$ and we define $\t\toua_n = \cup_i\t\toua_n^{(i)}$.

In order to check that the sequences on $n$ have zero limit in measure (that is, the limit $n\to\infty$ of the measure on the events of the sequence is zero), we just have to check this for $\toua_n$ and $\t\toua_n$. Let us first consider finite unions: the set of indices $i=1,2,\ldots,I$ is finite. Then, we have
\[
	\lim_{n\to\infty} P(\toua_n) \leq \lim_{n\to\infty} \sum_{i=1}^I P(\toua_n^{(i)}) =
	\sum_{i=1}^{I} \lim_{n\to\infty} P(\toua_n^{(i)}) = 0
\]
where the last equation is by strong continuity of $\tou^{(i)}$, and similarly for $\t\toua_n$. Hence, we have shown strong continuity for finite unions.

Let us consider infinite countable unions. Without loss of generality, and thanks to the result for finite unions and negations, we may consider $\tou^{(i)}$ to be mutually non-intersecting (just replace $\tou^{(i)}$ by $\tou^{(i)} \setminus \cup_{j=1}^{i-1} \tou^{(j)}$). Then, the series $\sum_{i=1}^\infty P(\tou^{(i)})_C$ must be convergent, so that $\lim_{I\to\infty} P(\cup_{i=I+1}^\infty \tou^{(i)})_C = \lim_{I\to\infty} \sum_{i=I+1}^\infty P(\tou^{(i)})_C = 0$: the sequence of events $\cup_{i=I+1}^\infty\tou^{(i)}$ form a decreasing sequence whose limit has zero measure. Consider
\[
	\lim_{n\to\infty} P(\toua_n) = \lim_{n\to\infty} P(\cup_i{\toua_n^{(i)}})_C.
\]
We may bound it as follows:
\beqa
	&& 0\leq \lim_{n\to\infty} P(\cup_i{\toua_n^{(i)}})_C \leq \lim_{n\to\infty} P(\cup_{i=1}^{I}{\toua_n^{(i)}} \cup \cup_{i=I+1}^\infty \tou^{(i)})_C
	\n && \leq \lim_{n\to\infty} \lt( P(\cup_{i=1}^{I}{\toua_n^{(i)}})_C + P(\cup_{i=I+1}^\infty \tou^{(i)})_C \rt)
	= P(\cup_{i=I+1}^\infty \tou^{(i)})_C \no
\eeqa
where in the last step we used strong continuity of the events and the fact that $I$ is finite. By the comments above, we may make the right hand side of the last equation as small as we want by choosing $I$ large enough. This shows that $\toua_n$ has zero limit in measure.

Following similar arguments as those above, the series $\sum_{i=1}^\infty P(\tou_n^{(i)})_C$ must be convergent, so that the sequence of events $\cup_{i=I+1}^\infty\tou_n^{(i)}$ for $I=1,2,3,\ldots$ form a decreasing sequence whose limit has zero measure. We need to be slightly more precise. Let us write temporarily $\toua_n^{(I)} \equiv \cup_{i=I+1}^\infty\tou_n^{(i)}$ (as well as $\toua^{(I)} \equiv \cup_{i=I+1}^\infty\tou^{(i)}$), and consider $\lim_{I\to\infty} P(\toua_{n(I)}^{(I)})_{C}$ for some $I$-dependent positive integers $n(I)$. If $n(I)$ is bounded for $I\in\N$, then the previous statement implies that this limit is zero. However, if it is not bounded, then there is no immediate implication. In this case, let us consider an infinite subsequence of $I$ for which $n(I)$ is non-decreasing and unbounded. If the limit of any such subsequence is zero, then the full limit $I\to\infty$ also is zero. We can bound (from above) the limit on a non-decreasing subsequence by $\lim_{I\to\infty} P(\cup_{n=n(I)}^\infty \toua_{n}^{(I)})_{C}$. The set $\cup_{n=n(I)}^\infty \toua_{n}^{(I)}$ is decreasing as $I$ increases, so that its limit exists. On the metric space of configurations, the distance between this set and $\cup_{n=n(I)}^{\infty}\toua^{(I)}_{g_n(C)}$ is a decreasing as $I$ increases, and this distance tends to 0. The latter set, $\cup_{n=n(I)}^{\infty}\toua^{(I)}_{g_n(C)}$, tends to $\toua^{(\infty)}_C$ possibly up to configurations where loops touch the boundary of $C$. Hence, the former set,$\cup_{n=n(I)}^\infty \toua_{n}^{(I)}$, tends to $\toua^{(\infty)}_C$ up to these configurations, and up to the closure under some smooth displacements (by $g_n^{-1}$) of the points of the loops that lie inside $C$. Since the measure on the set of configurations where loops touch the boundary of $C$ is zero, and since the measure of an event is unaffected by the closure under smooth deformations of the loops, the limit of the probability gives $P(\toua^{(\infty)})_C$, which is zero. This shows that $\lim_{I\to\infty} P(\cup_{i=I+1}^\infty \tou_{n(I)}^{(i)})_{C}=0$.

Consider
\[
	\lim_{n\to\infty} P(\t\toua_n)_C = \lim_{n\to\infty} P(\cup_i\t\toua_n^{(i)})_C.
\]
We may bound it as follows:
\beqa
	&& 0\leq \lim_{n\to\infty} P(\cup_i{\t\toua_n^{(i)}})_C\leq\lim_{n\to\infty} P(\cup_{i=1}^{I}{\t\toua_n^{(i)}}\cup\cup_{i=I+1}^\infty \tou_n^{(i)})_C
	\n && \leq \lim_{n\to\infty} \lt( P(\cup_{i=1}^{I}{\toua_n^{(i)}})_C + P(\cup_{i=I+1}^\infty \tou_n^{(i)})_C \rt)
	\leq P(\cup_{i=I+1}^\infty \tou_{n(I)}^{(i)})_C \no
\eeqa
where we choose for $n(I)$ the value of $n$ for which $P(\cup_{i=I+1}^\infty \tou_n^{(i)})_{C}$ is maximum. Then, by the result of the previous paragraph, we may make the right-hand side as small as we want by choosing $I$ large enough, so that $\t\toua_n$ has zero limit in measure. \eproof

Continuity is a somewhat straightforward concept, and expected to hold in many cases. In the present paper, we will only need continuity, but in the second part of this work, we will invoke differentiability. This is much harder to prove, although it is still expected in some cases. We do not have differentiability theorems yet, but we provide below a proof of Lipschitz continuity for some events of interest. For ordinary functions of one variable, this implies differentiability almost everywhere, hence it is much nearer to what we will need. Our proof is based, however, on one additional, but rather weak, assumption on the CLE measure, which will need independent proof.

First, let us make precise our notions of Lipschitz continuity.
\begin{defi}\label{deflcont}
An event $\tou$ is Lipschitz continuous at the simply connected domain $C$ with $\cl{C}$ not containing $\infty$, if
\beq
	\lim_{n\to\infty} \frc{P(\tou)_{g_n(C)} - P(\tou)_C}{\ep_n} <\infty,
\eeq
for any sequence of transformations $g_n$ conformal on $C$ with $\lim_{n\to\infty} g_n = \id$ on $\cl{C}$, and any sequence of $\ep_n>0$ such that $|g_n(z)-z|<\ep_n$ for all $z\in\b{C}$.
\end{defi}
Here, we use the notation $\lim_{n\to\infty} a_n <\infty$ (for $a_n\ge0$) with the meaning that there exists a finite $b>0$ such that for all $\delta>0$ there exists a $N$ such that $a_n-b<\delta$ for all $n>N$. That is, the limit itself may not exist, but the values obtained in the process of the taking limit must converge to a finite interval. The restriction to $\cl{C}$ not containing $\infty$ is for technical simplicity in the requirements on the sequence $\ep_n$; it can always be achieved by a conformal transformation.

Again, there is a useful intermediate step towards Lipschitz continuity.
\begin{defi}\label{defslcont}
An event $\tou$ is strongly Lipschitz continuous at the simply connected domain $C$ with $\cl{C}$ not containing $\infty$, if
\beq\label{eqslcont}
	\lim_{n\to\infty} \frc{P(\tou,\non g^{-1}_n\tou_{g_n(C)})_C}{\ep_n} <\infty
	\ \mbox{ and }\ \lim_{n\to\infty} \frc{P(g^{-1}_n \tou_{g_n(C)},\non \tou)_C}{\ep_n}<\infty
\eeq
for any sequence of transformations $g_n$ conformal on $C$ with $\lim_{n\to\infty} g_n = \id$ on $\cl{C}$, and any sequence of $\ep_n>0$ such that $|g_n(z)-z|<\ep_n$ for all $z\in\b{C}$.
\end{defi}
The relation between these concepts is as follows.
\begin{theorem}\label{theolcont}
An event that is strongly Lipschitz continuous at $C$ is Lipschitz continous at $C$.
\end{theorem}
\proof
Simply write
\beqa
	\lim_{n\to\infty} \frc{P(\tou)_{g_n(C)} - P(\tou)_C}{\ep_n} &=&
		\lim_{n\to\infty} \frc{P(\tou_n)_{C} - P(\tou)_C}{\ep_n} \n
		&=&  \lim_{n\to\infty} \frc{P(\tou_n,\non \tou)_{C} -
		P(\tou,\non \tou_n)_C}{\ep_n}
\eeqa
where $\tou_n \equiv g_n^{-1}\tou_{g_n(C)}$.
\eproof

We do not have a theorem that allows us to extend strong Lipschitz continuity from a set of events to its generated $\sigma$-algebra. The problem is in taking infinite series, as are involved in infinite countable unions. However, we may extend it to the algebra, involving only finite unions and negations:
\begin{theorem}\label{theolgener}
If all events in the set $\evs$ are strongly Lipschitz continuous at $C$, then all events in the algebra generated from $\evs$ are also strongly Lipschitz continuous at $C$.
\end{theorem}
\proof We adjoin the trivial event $\conf_\C$ (which is strongly Lipschitz continuous) to $\evs$ if it is not there. The negation of events in $\evs$ is strongly Lipschitz continuous from the definition. Let us consider the union of two events $\tou$ and $\tou'$ in $\evs$, and write $\tou_n = g^{-1}_n \tou_{g_n(C)}$, $\tou_n' = g^{-1}_n \tou_{g_n(C)}'$. We have
\[
	P(\tou\cup \tou',\non(\tou_n\cup\tou_n'))_C \leq P((\tou\cap \non\tou_n) \cup (\tou'\cap \non\tou_n'))_C \leq
	P(\tou\cap \non\tou_n)_C + P(\tou'\cap \non\tou_n')_C
\]
and similarly for $P(\non(\tou\cup\tou'),\tou_n\cup\tou_n')_C$. \eproof

\subsection{Events of interest}\label{subsectevents}

\begin{figure}
\bc
\includegraphics[width=7cm,height=7cm]{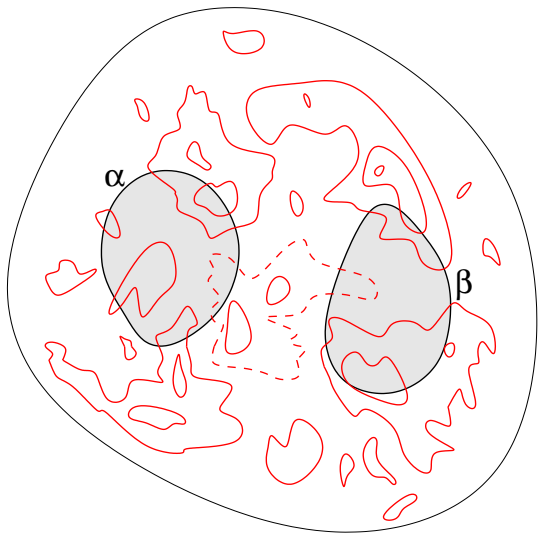}
\ec
\caption{The event $\ev(\alpha,\beta)$ on the configuration depicted in figure \ref{figCLE}. The dashed CLE loop breaks the conditions of the event. The shaded areas are the natural domains associated to the loops $\alpha$ and $\beta$.}
\label{figevents}
\end{figure}

We now introduce the events in $\evs(\conf_\C)$ that will be of most interest for the present work. They are characterised by two disjoint simple loops $\alpha,\beta$, and will be denoted by $\ev(\alpha,\beta)$. They are defined by the requirement that there is no CLE loop that intersects both $\alpha$ and $\beta$. It will be convenient to associate to each of $\alpha$ and $\beta$ a different simply connected domain, bounded by $\alpha$ or $\beta$, which we will call the {\em natural domain} associated to $\alpha$ or $\beta$. The natural domain associated to $\alpha$ is the simply connected component of $\C\setminus \alpha$ that does not contain $\beta$, and {\em vice versa} for the natural domain associated to $\beta$. Certainly, then, the requirement defining the event $\ev(\alpha,\beta)$ is equivalent to imposing that no loop intersects both closures of the natural domains associated to $\alpha$ and $\beta$ (not taking the closure of the domains would give the same event in measure). See figure \ref{figevents}. In general, these events have non-zero probability on any simply connected domain $C$ (no matter what $\alpha\cup\beta$ is). However, we will mostly restrict our attention to $\alpha\cup\beta\subset C$ when we consider probabilities on $C$.

The idea behind these events is that they produce a ``separation'' between the natural domains associated to $\alpha$ and $\beta$, by forbidding that these two domains be affected by a common CLE loop. Of course, the loop configurations in the two domains do not become independent, because of chains of mutually influencing loops connecting them. However, the way by which we will obtain a CLE probability function on annular domains will be by taking $\alpha$ infinitely near to $\beta$, with an appropriate re-normalisation; then, they indeed become independent. Since according to the ideas of \cite{DRC06} probabilities on annular domains are related to the stress-energy tensor, it is also in this way that we will define the stress-energy tensor in the second part of this work, essentially choosing a different re-normalisation.

We should remark that for many aspects of this work, we could have used, instead, events defined by the condition that at least one loop separates $\alpha$ from $\beta$; this is again in the spirit of asking for a ``disconnection'' between the natural domains associated to $\alpha$ and $\beta$. Many of the theorems below hold for these events. However, certain theorems rely on the particular properties of the events $\ev(\alpha,\beta)$ introduced above, hence for simplicity we only discuss these events.

\begin{figure}
\bc
\includegraphics[width=7cm,height=7cm]{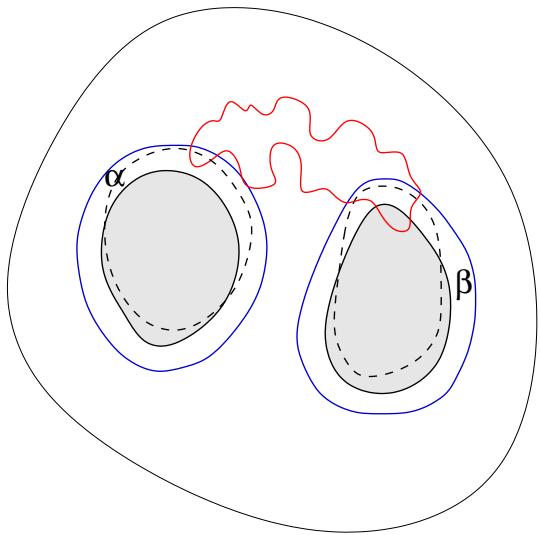}
\ec
\caption{The loops $\alpha$ and $\beta$ of the event (full black line) along with their natural domains (shaded area), their conformal transforms under $g_n$ (dashed black line) and the loops $\alpha_n$ and $\beta_n$ (blue line). A CLE loop (red) that intersects both conformal transforms but not $\alpha$, or not $\beta$, also intersects both $\alpha_n$ and $\beta_n$ but not $\alpha$, or not $\beta$, so satisfies the conditions of $\toua_n$.}
\label{figproofcont}
\end{figure}

We first prove strong continuity for $\ev(\alpha,\beta)$. This theorem will turn out to be quite useful for proving continuity for more general events. It is a consequence of very few properties of CLE: essentially only that taking open or closed domains does not change the measure of events. We do not need explicitly any of the three defining axioms of CLE (except for conformal invariance, indirectly in the definitions of continuity and strong continuity; but this does not play an essential r\^ole).
\begin{theorem}\label{theoscev}
The event $\ev(\alpha,\beta)$ is strongly continuous at $C$ for any $C$ containing $\alpha\cup \beta$.
\end{theorem}
\proof We will construct two decreasing covering sequences of events $\toua_n$ and $\t\toua_n$, that cover $\tou\cap\non g^{-1}_n\tou_{g_n(C)}$ and $g^{-1}_n \tou_{g_n(C)}\cap\non \tou$ respectively, for $\tou = \ev(\alpha,\beta)$, and we will prove that the limits $\cap_n\toua_n$ and $\cap_n\t\toua_n$ are events of zero measure. Let us start with the sequence $\toua_n$. For a fixed $N$, consider a $\delta>0$ such that the distance between $g_n(z)$ and $z$ is smaller than $\delta$ (recall the notion of distance in appendix \ref{notations}) for all $n>N$ and for all $z$ on the loop $\alpha$. Let us construct the loop $\alpha_N$ in the exterior of the natural domain bounded by $\alpha$, such that all points of $\alpha_N$ are a distance $\delta$ away from $\alpha$. Similarly, for the same $N$, consider a (possibly different) $\delta$ associated to $\beta$, and let us construct the loop $\beta_N$ in the exterior of the natural domain bounded by $\beta$. Consider the event $\toua_N$ that at least one loop intersects $\alpha_N$ and $\beta_N$, but either it doesn't intersect $\alpha$, or it doesn't intersect $\beta$ (or both). See figure \ref{figproofcont}.
Clearly, $\toua_N$ forms, over $N$, a covering sequence, and it is possible to choose the $\delta$'s decreasing, so that it is a decreasing sequence. In particular, it is possible to choose $\lim_{N\to\infty} \delta = 0$, which we do. Then, $\cap_n\toua_n$ is the empty event, which has measure zero. This completes the first part of the proof. For the sequence $\t\toua_n$, we similarly construct loops $\alpha_N$ and $\beta_N$, but now they must be inside the natural domains. The event $\t\toua_N$ we consider is that at least one loop intersects $\alpha$ and $\beta$, but either it doesn't intersect $\alpha_N$, or it doesn't intersect $\beta_N$ (or both). Again, we can make it a decreasing covering sequence. The event $\cap_n\t\toua_n$ is now that at least one loop intersects $\alpha$ and $\beta$ without intersecting at least one of the natural domains, but this is the same in measure as the event with the closure of the natural domains, which is the empty event.
\eproof

We now investigate strong Lipschitz continuity. This is not as straightforward a consequence of the measure as strong continuity above is. In particular, it needs in an essential way conformal invariance of the CLE measure. It also needs a more stringent finiteness property, related to, but stronger than, the finiteness II property: that in the CLE measure, the average of the number of loops of extent at least $d$ is less than infinity for any $0<d<1$. That is, we require the probability of finding exactly $n$ loops of extent at least $d$ to decay fast enough as $n\to\infty$. We do expect this to be true for $8/3<\kappa<4$; for instance, as $\kappa\to8/3$, the ``density'' of loops decreases to zero. But in order to admit all possibilities, we will say that we choose values of $\kappa$, if any, where this property holds. For simplicity, we will say that we choose values of $\kappa$ with {\em finite averages}.

Interestingly, though, besides conformal invariance (and the property of finite averages), we do not need other of the defining properties of the CLE measure. Hence, Lipschitz continuity may well hold in more general measures than that of CLE.

In order to express the theorem in its generality, we first introduce a concept of smoothness of a loop $\alpha$ in some domain $C$. Let us define this concept in $\uD$, and use conformal transport to define it in other simply connected domains. In the theorem below, we say that a loop $\alpha$ in $\uD$ is smooth if we can choose a finite set of open arcs of $\alpha$ that cover $\alpha$, with the property that there exists a $t_0>0$, and for each arc $\gamma$ there exists a family $g_t^\gamma,\,t\in[-t_0,t_0]$ of conformal transformations preserving $\uD$, such that $g_t^\gamma(\gamma)\cap g_{t'}^\gamma(\gamma)=\emptyset\;\forall t,t' \in [-t_0,t_0],\,t\neq t'$. That is, we must be able to divide $\alpha$ into a finite set of open arcs, where each of these arcs sweeps just the interior or just the exterior of $\alpha$ under an approprite $\uD$-preserving conformal transformation.

\begin{theorem}\label{theoslcev}
For $\alpha$ and $\beta$ smooth as defined above, the event $\ev(\alpha,\beta)$ is strongly Lipschitz continuous at $C$ for any $C$ containing $\alpha\cup \beta$ for any value of $\kappa$ with finite averages.
\end{theorem}
The proof goes by showing that the events $\toua_n$ constructed in the proof of theorem \ref{theoscev} in fact have measures that vanish proportionally to $\delta$ as $\delta\to0$ (that is, as $n\to\infty$). The way this is done is by showing such a vanishing for other events over which we have better control thanks to conformal invariance, and which we can use to cover $\toua_n$.

\proof 
We restrict ourselves to $\uD$ by conformal invariance -- this just affects the shapes of $\alpha$ and $\beta$, but keeps smoothness and the angles of the possible corners.

\begin{figure}
\bc
\includegraphics[width=7cm,height=7cm]{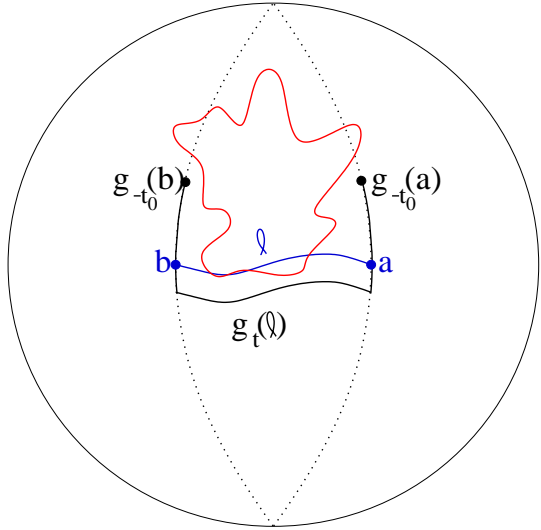}
\ec
\caption{In blue is the curve segment $\ell$, and in full black, its conformal transform $g_t(\ell)$ and the set of conformal transforms of its boundary points $a$ and $b$, $\{g_{t'}(\p\ell)\,|\,t'\in[-t_0,t]\}$. The dotted lines indicate the set of points given by $g_{t'}(\p\ell)$ for $t'\in\R$. There must be at least one loop that crosses the blue segment without crossing the black segments, like the CLE loop in red.}
\label{figprooflcont}
\end{figure}

We start by considering, instead of $\alpha$ and $\beta$, a continuous, simple, finite curve segment $\ell$ in $\uD$. We choose it such that under a $\uD$-preserving abelian group of transformations $g_t$ (for instance, $g_t(z) = (z\cosh t-i\sinh t)/(iz\sinh t+\cosh t)$) for $t\in\R$ (with $g_t\circ g_{t'} = g_{t+t'}$), we have $g_t(\ell) \cap g_{t'}(\ell) = \emptyset$ for any pair of $t$ and $t'$, with $t\neq t'$, lying in an open interval containing 0. Then we show that the probability $p_t$ that there is at least one loop, of extent at least $d$, that intersects $\ell$ but does not intersect $g_t(\ell)\cup \{g_{t'}(\p \ell)\,|\,t'\in[-t_0,t]\}$, satisfies $\lim_{t\to 0^+}p_t/t <\infty$ for any $0<d<1$ and any $t_0>0$. See figure \ref{figprooflcont}.

\begin{figure}
\bc
\includegraphics[width=14cm,height=7cm]{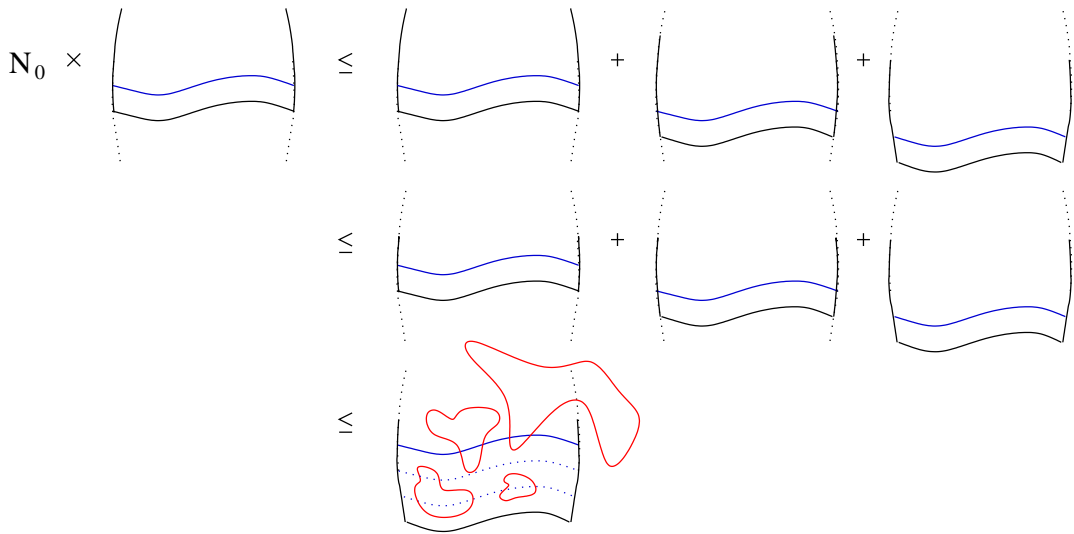}
\ec
\caption{A pictorial representation of the steps of the proof, with $N_0=3$. The drawings without loops are probability that at least one loop intersects the blue segment without intersecting the black segments. The last drawing, with loops, is the average of the number of loops that intersect the region bounded by the blue and black segments.}
\label{figprooflcont2}
\end{figure}

The proof goes as follows. See figure \ref{figprooflcont2} for a pictorial representation of the steps. For a given integer $N\ge1$, consider the probabilities $p_n(d')$ that at least one loop of extent at least $d'$ intersects $g_{n/N}(\ell)$ but not $g_{(n+1)/N}(\ell)\cup \{g_{t'}(\p \ell)\,|\,t'\in[n/N-t_0,(n+1)/N]\}$, for $n=0,1,2,\ldots,N_0-1$ with $N_0 = [Nt_0]+1$ (here, $[x]$ is the integer part of $x$). Note that $p_0(d) = p_{1/N}$. The conformal transformation $g_{-n/N}$ of the conditions of intersections on the loops for $p_n(d')$ gives the conditions for $p_0(d')$, for any $n$. Hence, we can hope to use conformal invariance to relate them. The extent is not conformally invariant, but for any $d$ with $0<d<1$, there is a $d'$ with $0<d'<1$ such that the extent of $g_{n/N}(\gamma)$ is at least $d'$ for any loop $\gamma$ of extent at least $d$ and for any $n=0,1,2,\ldots,N-1$. Choosing such a $d'$, we then have $p_n(d')\ge p_0(d)$. Hence, $\sum_{n=0}^{N_0-1} p_n(d') \ge N_0 p_{1/N}$. This is the first step. Let us consider the probabilities $\t{p}_n(d')$ that at least one loop of extent at least $d'$ intersects $g_{n/N}(\ell)$ but not $g_{(n+1)/N}(\ell)\cup \{g_{t'}(\p \ell)\,|\,t'\in[0,(n+1)/N]\}$. That is, $\t{p}_n(d')$ has, for the set with non-intersecting condition, a set smaller than or equal to that of $p_n(d')$ for all $n$. Hence we have $\t{p}_n(d') \ge p_n(d')$ for all $n$. This is the second step. Finally, let us consider the random variable $\eta$ which gives the number of loops of extent at least $d'$ that intersect the region bounded by the segments $\ell$ and $g_{N_0/N}(\ell) \cup \{g_{t'}(\p \ell) \,|\, t'\in[0,N_0/N]\}$, but that do not intersect $g_{N_0/N}(\ell) \cup \{g_{t'}(\p \ell) \,|\, t'\in[0,N_0/N]\}$. Thanks to the assumption of finite averages, we have for the CLE average $\bra \eta\ket<\infty$ and $\lim_{N\to\infty} \bra\eta\ket < \infty$. Using $\t\pres_n(d')$ for the characteristic function associated with $\t{p}_n(d')$, i.e. $\t{p}_n(d') = \bra \t\pres_n(d')\ket$, in every configuration, the sum $\sum_{n=0}^{N_0-1} \t\pres_n(d')$ is less than or equal to $\eta$ (every loop is counted at most once). Hence, $\sum_{n=0}^{N_0-1} \t{p}_n(d')\leq \bra \eta\ket$. This is the third step. Combining everything, $\lim_{N\to\infty} N_0p_{1/N} = t_0(\lim_{N\to\infty} Np_{1/N})  < \infty$ which shows the assertion.

\begin{figure}
\bc
\includegraphics[width=7cm,height=3cm]{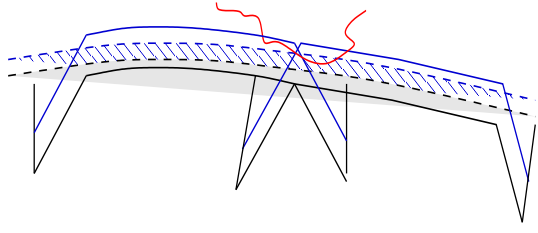}
\ec
\caption{A patch corresponds to a structure topologically like that of the left-hand side of figure \ref{figprooflcont2}. In the present figure, dashed black lines are $\alpha$, dashed blue lines are $\alpha_n$, and black and blue lines are patches. On the left, part of a covering of an arc, with two patches. A loop that intersects $\alpha_n$ and $\beta_n$ ($\beta_n$ is further away, not shown here) but not $\alpha$, satisfies the conditions of at least one patch (for some finite extent).}
\label{figprooflcont3}
\end{figure}

Let us now consider the events $\toua_n$ of theorem \ref{theoscev}, associated to $\ev(\alpha,\beta)$, and the loops $\alpha_n$ and $\beta_n$ related to $\toua_n$. We need to cover the ``rim'' formed by $\alpha$ and $\alpha_n$, and that formed by $\beta$ and $\beta_n$ (see again figure \ref{figprooflcont}). Let us consider a choice of $\ell$, of direction of $g_t$, and of $t$, such that there is a part of $\ell$ that lies outside the natural domain associated to $\alpha_n$, yet all of $g_t(\ell)\cup \{g_{t'}(\p \ell)\,|\,t'\in[-t_0,t]\}$ lies inside the natural domain associated to $\alpha$. We will call this a ``patch'' of the rim $\alpha,\alpha_n$. See figure \ref{figprooflcont3}. For $\alpha$ and $\beta$ smooth as defined above, it is always possible to find a finite number $m$ of patches such that both rims are completely covered by patches, for all $\toua_n$. Denote by $\tou_n^{(i)},\,i=1,\ldots,m$ the events corresponding to these patches. We have
\[
	P(\toua_n)_\uD\leq P(\cup_{i=1}^m \tou_n^{(i)})_\uD \leq \sum_{i=1}^m P(\tou_n^{(i)})_\uD.
\]
Since we can always choose $t$ of the order of $\delta$ for all $i$, this shows the first inequation of \ref{eqslcont}. For the second inequation, we take the events $\toua_n$ associated to $\non\ev(\alpha,\beta)$, but they are exactly of the same form, so it holds as well.
\eproof

Three comments are in order. First, the proof can be used more generally to show (an extended version of) continuity of $\ev(\alpha,\beta)$ under {\em any} deformations of $\alpha$ and $\beta$. Lipschitz continuity in definition \ref{deflcont} uses only conformal transformations on $C$, because this is sufficient for our purposes and allows a general definition without reference to the particulars of the CLE events. Second, using small deformations of $\alpha$ and $\beta$ instead of conformal transformations of events, it is straightforward from the result above to show strong Lipschitz continuity for {\em restricted} probabilities, restricted to any fixed event of non-zero probability. Third, the techniques of the proof above can be applied as well to many other events of similar type, which might not even be in the algebra generated by $\ev(\alpha,\beta)$.

\subsection{Support}\label{subsectsupport}

The concept of ``locality'' plays a fundamental r\^ole in quantum field theory. Essentially, it is related to how much of the space a certain object covers or ``feels''. Similarly, in the context of our construction in CLE it will be crucial to have a concept of support of an event in CLE. Essentially, a support is a set in $\C$ such that if any loop or any actual domain of restriction separates this support from the rest, then the event is only determined by the loops in that part of the configuration.
\begin{defi}\label{supp}
A support $\supp(\tou)$ of an event $\tou\in\ev(\conf_\C)$ is a closed subset of $\C$, with the following properties:
\begin{enumerate}
\item \label{firstpoint} $\tou_C\in\sgm_C$ for any simply connected domain $C$ that includes $\supp(\tou)$.
\item In instances of a CLE on any simply connected domain $C\subset \C$, if $\supp(\tou)$ is surrounded by a loop $\gamma$ (in particular, they do not intersect), the evaluation of the event is that obtained from the configuration inside the loop. More precisely, consider $\sam_C$ and suppose a configuration $x_j^C$ contains a loop $\gamma$ that surrounds $\supp(\tou)$, i.e. $\supp(\tou) \subset C_\gamma$. Then
\beq\label{support1}
    \pres(x_j^C,\tou) = \pres(\inter_{C_\gamma}(x_j^C),\tou).
\eeq
\item The configuration inside any loop that does not surround neither intersect $\supp(\tou)$ does not affect the evaluation of the event. Consider $\sam_C$, and for a configuration $x_j^C$, select a set of loops $\Gamma$, in the exterior of each other, that do not have any part of $\supp(\tou)$ in their interior, and that do not intersect $\supp(\tou)$:
\[
	\supp(\tou) \subset C\setminus \overline{\cup_{\gamma\in\Gamma} C_\gamma}
\]
(there is always such a set of loops). Then
\beq\label{support2}
    \pres(x_j^C,\tou) = \pres(\exter_{\cup_{\gamma\in\Gamma} C_\gamma}(x_j^C),\tou).
\eeq
\item If $\supp(\tou)$ is inside a simply connected component of an actual domain of restriction, the evaluation of the event $\tou$ is that obtained from the configuration inside this domain. Consider $\sam_C$ and suppose that in some configuration $x_j^C$, we have $\supp(\tou) \subset A$, where $A$ is a simply connected component of an actual domain of restriction of $x_j^C$. Then
\beq\label{support3}
    \pres(x_j^C,\tou) = \pres(\inter_{A}(x_j^C),\tou).
\eeq
\item If $\supp(\tou)$ is outside a simply connected component of an actual domain of restriction, the evaluation of the event $\tou$ is not affected by the configuration inside this component. Consider $\sam_C$, and suppose that in some configuration $x_j^C$, there is a set $\Omega$ of simply connected components of an actual domain of restriction, whose closures do not intersect $\supp(\tou)$. Then
\beq\label{support4}
    \pres(x_j^C,\tou) = \pres(\exter_{\cup_{A\in\Omega} A}(x_j^C),\tou).
\eeq
\end{enumerate}
\end{defi}
It will also be convenient to introduce the notion of a non-zero supported event: we will say that an event $\tou$ is non-zero on its support if $P(\tou)_C>0$ for any simply connected domain $C$ that includes $\supp(\tou)$.

The support of an event is in general not unique:
\begin{corol}\label{corolaugmsupp}
If $A$ is a support of $\tou\in\ev(\conf_\C)$, then any closed set $B$ such that $A\subseteq B$ is also a support.
\end{corol}
\proof A straightforward inspection of the five points in the definition \ref{supp} shows that this is indeed the case.
\eproof

In general, the support of a conjunction or a union of events is just the union of their supports, which by the previous corollary is a good support for both events:
\begin{corol}\label{corolconj}
For two events $\tou$ and $\tou'$ possessing a support, we may take $\supp(\tou\cap\tou') = \supp(\tou\cup\tou') = \supp(\tou)\cup \supp(\tou')$.
\end{corol}
\proof Use corollary \ref{corolaugmsupp}, the properties of $\sigma$-algebras, and $\pres(x,\tou\cap\tou') = \pres(x,\tou)\pres(x,\tou')$ and $\pres(x,\tou\cup\tou') = 1-(1-\pres(x,\tou))(1-\pres(x,\tou'))$. \eproof

A conformal transformation of an event $\tou$, conformal on $\supp(\tou)$, should have a support that is the conformal transform of $\supp(\tou)$. There is a subtlety, as the domain where the transformation is conformal may not be $\C$: we need to restrict the event to $\conf_A$ for a domain $A$ where the transformation is conformal. Then, in the definition of the support for the transformed event under $g$, we must consider only simply connected domains of definition included inside $g(A)$. With this restriction, this will be called a $g(A)$-reduced support. In the case where $A=\C$, this is indeed a support as defined above. We will not make much use of reduced supports, but we note that corollaries \ref{corolaugmsupp} and \ref{corolconj} have natural analogues for reduced supports. In general, we have:
\begin{corol}\label{coroltranssupp}
If $\tou$ has a support and $\supp(\tou)\subset A$, then the $g(A)$-reduced support of $g(\tou_A)$ for $g$ conformal on the domain $A$ is $g(\supp(\tou))$. If $g$ is a global conformal transformation, then this is a true support of $g(\tou)$.
\end{corol}
\proof This is an immediate consequence of the definition of support. \eproof

The event $\conf_\C$, without any conditions, possesses a support, which can be taken as the empty set. This event is in fact non-zero on that support. The event $\emptyset$ also possesses the same support, but it is zero on it, as well as on any other support. For the event $\ev(\alpha,\beta)$, the support can be taken as $\alpha\cup\beta$. In many cases, it is quite straightforward to verify if an event possesses a support and to find examples.

The properties of a support will be at the basis of many of the proofs and arguments in the following sections. They give rise to quite strong statements when put in conjunction with the fundamental properties of the CLE measure. In conjunction with the nesting property as expressed in subsection \ref{ssectrandom}, property (\ref{support1}) means that, for instance, with $M$ a set of integers such that $x_j^C$ contains an appropriately chosen (subsection \ref{ssectrandom}) loop that surrounds $\supp(\tou)$ for $j\in M$, and with $\gamma_j\in x_j^C$ such a loop, we can write from (\ref{nestprop})
\beq\label{relation1}
	P(\tou)_C = \lim_{N\to\infty} N^{-1}\sum_{j=1}^N
	\lt\{\ba{ll} \pres(x_j^C,\tou) & (j\not\in M) \\
	P(\tou)_{C_{\gamma_j}} & (j\in M). \ea\rt.
\eeq
On the other hand, we may combine properties (\ref{support1}) and (\ref{support2}) when we have two events in conjunction, using $\pres(x,\tou\cap \t\tou) = \pres(x,\tou)\pres(x,\t\tou)$. If $\tou$ and $\t\tou$ are supported away from each other (i.e.\ have disjoint supports), and for an appropriate set $M$ such that the configuration $x_j^C$ has an appropriately chosen loop $\gamma_j$ that contains $\supp(\tou)$ and separates it from $\supp(\t\tou)$ for all $j\in M$, we have
\beq\label{relation3}
	P(\tou,\t\tou)_C = \lim_{N\to\infty} N^{-1}\sum_{j=1}^N
	\lt\{\ba{ll} \pres(x_j^C,\tou) \pres(x_j^C,\t\tou) & (j\not\in M) \\
	P(\tou)_{C_{\gamma_j}} \pres(\exter_{C_{\gamma_j}}(x_j^C),\t\tou)  & (j\in M). \ea\rt.
\eeq
Using (\ref{support2}) again, this can be written
\beq\label{relation4}
	P(\tou,\t\tou)_C = \lim_{N\to\infty} N^{-1}\sum_{j=1}^N
	\pres(x_j^C,\t\tou) \cdot \lt\{\ba{ll} \pres(x_j^C,\tou) & (j\not\in M) \\
	P(\tou)_{C_{\gamma_j}}  & (j\in M). \ea\rt.
\eeq

We may also make use of restricted re-randomisation procedures for restricted probabilities in formulas (\ref{relation1})-(\ref{relation4}). In particular, with $\tou'$ non-zero on its support, we have, instead of (\ref{relation4}),
\beq\label{relation4rest}
	P(\tou,\t\tou|\tou')_C = \lim_{N\to\infty} N^{-1}\sum_{j\in I_{\tou'}^{(N)}}
	\pres(x_j^C,\t\tou) \cdot \lt\{\ba{ll} \pres(x_j^C,\tou) & (j\not\in M) \\
	P(\tou|\tou')_{C_{\gamma_j}}  & (j\in M). \ea\rt.
\eeq
where now $M$ is a set such that the configuration $x_j^C$ has a loop $\gamma_j$ that contains $\supp(\tou)\cup \supp(\tou')$ and separates it from $\supp(\t\tou)$ for all $j\in M$, and $I_{\tou'}^{(N)}$ is the set of the $N$ smallest integers $j$ such that $x_j^C$ satisfies the conditions of $\tou'$.

Similar formulas can also be obtained using (\ref{support3}) and (\ref{support4}), related to conformal restriction, instead of (\ref{support1}) and (\ref{support2}), related to nesting.

Our main theorem associated to the notion of support is that of continuity of supported events. The proof uses continuity of all events in the $\sigma$-algebra generated by $\ev(\alpha,\beta)$. This continuity is a consequence of theorems \ref{theoscev}, \ref{theogener} and \ref{theocont}. The proof also uses nesting and conformal restriction in conjunction with the properties of support.

Let us now define a family of measures on certain simple loops in a simply connected domain, induced from the CLE measure and the events $\ev(\alpha,\beta)$. This construction will be of great use later on as well.

\begin{figure}
\bc
\includegraphics[width=12cm,height=5cm]{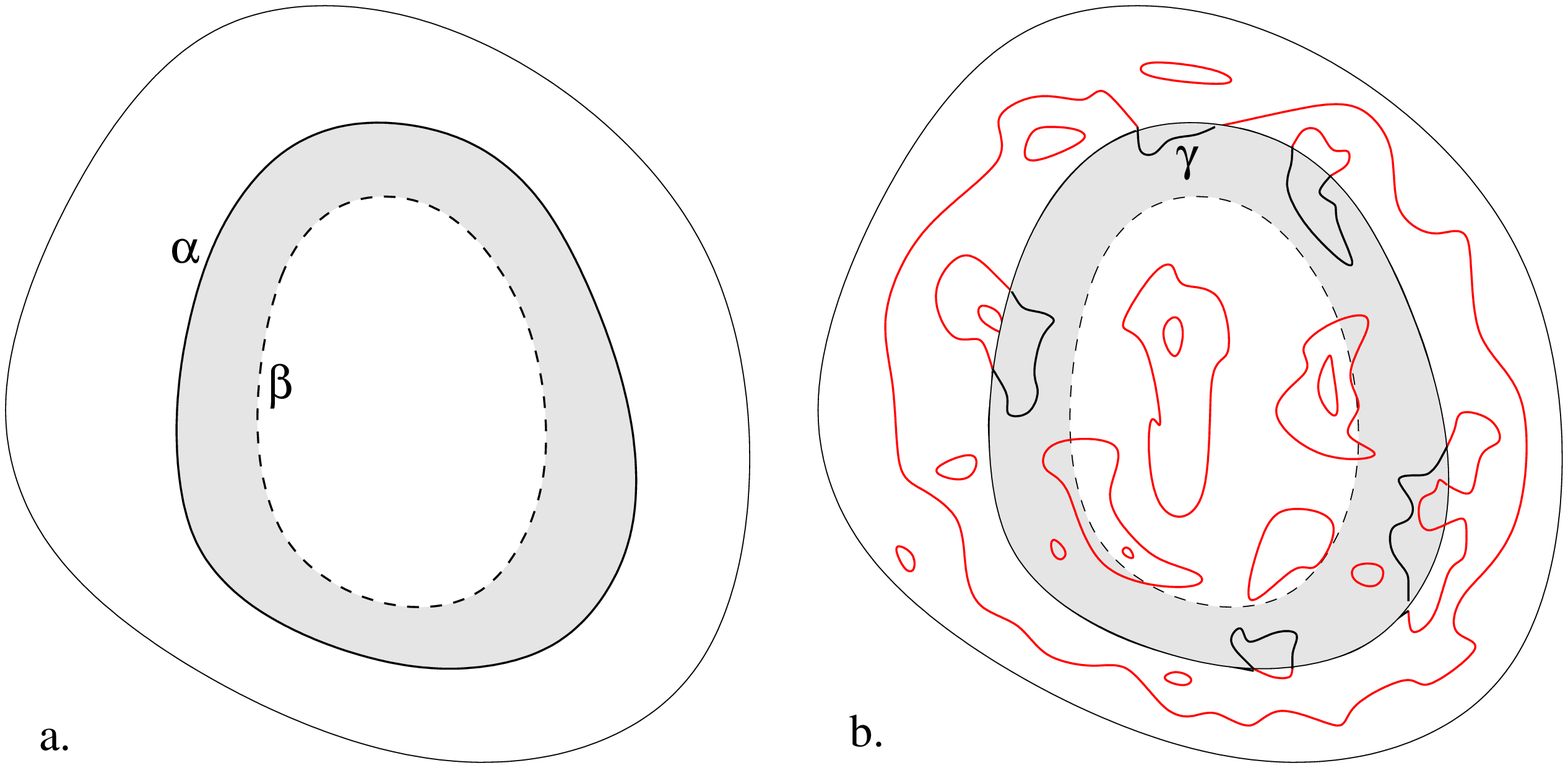}
\ec
\caption{Construction of the loop inside the annulus delimited by $\alpha$ and $\beta$. a. The annular region. b. The random loop constructed from a given CLE configuration. In the case shown, there is exactly one ``surrounding'' loop (that surrounds $\beta$ and intersects $\alpha$). The loop $\gamma$ is obtained by looking at the domain inside the surrounding loop and by constructing the component of the actual domain of restriction that contains $\beta$. Equivalently, $\gamma$ bounds the component that contains $\beta$ of the intersection of: the interior of $\alpha$, the interior of any loop that intersects $\alpha$ and surrounds $\beta$, and the exterior of any loop that intersects $\alpha$ but does not surround $\beta$. The loop $\gamma$ always has points in common with $\alpha$, but no arcs (because almost every point is surrounded by a loop), contrary to this crude representation.}
\label{figtheloop}
\end{figure}

We start by constructing the loop from any given CLE configuration. Consider CLE on $C$, and the case where the natural domain associated to $\alpha$ is {\em not included} inside $C$: it has a part outside $C$. In this case, we have $\beta\subset C_\alpha$, and the region between $\alpha$ and $\beta$ looks like an annulus in $C$ (see figure \ref{figtheloop} a.). Let us consider CLE restricted on the event $\ev(\alpha,\beta)$, and in this, consider conformal restriction based on $\cl{C_\alpha}$. Then there are two possibilities for the actual domain of restriction: whether it completely includes $\beta$, or it does not intersect $\beta$ at all. The case where it does not intersect $\beta$ occurs if and only if a loop intersects $\alpha$ and surrounds $\beta$; let us call this a ``surrounding'' loop. We construct a simple loop $\gamma$ inside the annular region delimited by $\alpha$ and $\beta$ as follows (it always intersects $\alpha$ but never $\beta$). If there are surrounding loops, we look at the last one, which does not surround any other surrounding loop. In the domain bounded by this loop, we consider the domains involved in conformal restriction based on the part of $\cl{C_\alpha}$ that is inside the loop. The loop $\gamma$ is just the boundary of the component of the resulting actual domain of restriction that contains $\beta$ (see figure \ref{figtheloop} b.).

The construction above is purely configurational, without reference to the CLE measure. Consider the space $\conf(\alpha,\beta)$ of all loops completely lying on, and ``going around'', the annular region $A_{\alpha,\beta}$ delimited by $\alpha$ and $\beta$ (close at the $\alpha$ boundary, and open at the $\beta$ boundary). It is clear that the construction above induces a mapping between CLE events restricted to $\ev(\alpha,\beta)$, and subsets of $\conf(\alpha,\beta)$ (that is, events on the loop $\gamma$ constructed above). This mapping preserves set operations. Hence, the CLE $\sigma$-algebra induces a $\sigma$-algebra on $\conf(\alpha,\beta)$, and the CLE measure induces a finite measure on this $\sigma$-algebra. We will denote the latter measure by $\omega_{C;\alpha,\beta}$. Consider the events $e_{\beta'}$ that the loop $\gamma\in\conf(\alpha,\beta)$ lies on the generically smaller annular region $A_{\alpha,\beta'}$, for any $\beta'$ lying on, and going around, the closure of $A_{\alpha,\beta}$. That is, these are the events that the random loop does not go further than $\beta'$. The $\sigma$-algebra on which $\omega_{C;\alpha,\beta}$ is defined can be taken to be the completion\footnote{The completion of a $\sigma$-algebra with respect to a given measure is essentially the addition of events of zero measure, that are subsets of events already present \cite{H50}.} of the $\sigma$-algebra generated by $e_{\beta'}$ for all $\beta'$. The mapping from CLE events is, in the cases of the generating events, $\ev(\alpha,\beta')\mapsto e_{\beta'}$. The associated measure is $\omega_{C;\alpha,\beta}(e_{\beta'}) = P(\ev(\alpha,\beta')|\ev(\alpha,\beta))_C$. The expectation $\int f(\gamma) d\omega_{C;\alpha,\beta}(\gamma)$ of a measurable function $f$ will sometimes be denoted $\bra f\ket_{C,\alpha,\beta}$.

First, we show that expectations of bounded nonnegative measurable functions are, as functions of the simply connected domain $C$, continuous:
\begin{theorem}\label{theocontexp}
For any bounded, nonnegative function $f$ on $S(\alpha,\beta)$, measurable with respect to  $\omega_{C;\alpha,\beta}$, the function $F: C\mapsto \bra f\ket_{C,\alpha,\beta}$ is continuous at any simply connected domain $C$ containing $\alpha$ and $\beta$. Moreover, in the notation of definition \ref{defcontgen}, for all $\delta>0$, there exists a $N$ that depends on $f$ only through its maximum $f_{max}$, such that for all $n>N$, $|\bra f\ket_{g_n(C),\alpha,\beta} - \bra f\ket_{C,\alpha,\beta}| <\delta$.
\end{theorem}
Concerning the last part, we will say that {\em the continuity statement only depends on $f_{max}$}. In symbols, this is:
\beq
	\forall \delta>0:\;\exists N\;|\;\forall f | {\rm max}(f)\leq f_{max},\,\forall n>N:\;
	|\bra f\ket_{g_n(C),\alpha,\beta} - \bra f\ket_{C,\alpha,\beta}| <\delta.
\eeq
\proof By theorems \ref{theoscev}, \ref{theogener} and \ref{theocont}, $\omega_{C;\alpha,\beta}$ is continuous, as a function of $C$, on any event in its $\sigma$-algebra, for $\alpha$ and $\beta$ in $C$ (satisfying the conditions above). Let us consider the signed measure $\nu_n  = \omega_{g_n(C);\alpha,\beta} - \omega_{C;\alpha,\beta}$.  By the Jordan decomposition, we can consider the upper and lower variations $\nu_n^\pm$ (which are both positive measures on the same $\sigma$-algebra) and write $\nu_n = \nu_n^+ - \nu_n^-$. Then,
\beqa
	\lt|\bra f\ket_{g_n(C),\alpha,\beta}-\bra f\ket_{C,\alpha,\beta}\rt| &=&
		\lt|\int f(\gamma) d\nu_n^+(\gamma) - \int f(\gamma) d\nu_n^-(\gamma)\rt| \n && \leq
		\lt|\int f(\gamma) d\nu_n^+(\gamma)\rt| + \lt|\int f(\gamma) d\nu_n^-(\gamma)\rt| \leq f_{max} (\nu_n^+(\conf) + \nu_n^-(\conf)) \no
\eeqa
where $\conf$ is the sure event (the whole space of loops $\gamma$). But by the Hahn decomposition we have $\nu_n^\pm(E) = \nu_n(E\cap A_{\pm})$ for $A_\pm$ some subsets of $\conf$, where $E\cap A_\pm$ are in the $\sigma$-algebra if $E$ is. Then, by continuity, which follows from theorems \ref{theoscev}, \ref{theogener} and \ref{theocont}, from $\omega_{C;\alpha,\beta}(e_{\beta'}) = P(\ev(\alpha,\beta')|\ev(\alpha,\beta))_C$ and from the fact that $e_{\beta'}$ form a set of generating events, we have $\lim_{n\to\infty} \nu_n^\pm(E) = 0$ for any event $E$ in the $\sigma$-algebra. \eproof

Note that the Jordan and Hahn decompositions \cite{H50} used in this proof are in the same spirit as that of the decomposition used to relate continuity to strong continuity, in the proof of theorem \ref{theocont}, for instance.

Next, we show that CLE probabilities restricted to $\ev(\alpha,\beta)$ can be expressed using the measure $\omega(C;\alpha,\beta)$:
\begin{theorem}\label{theoCLEomega}
For $C$ a simply connected domain and $\alpha$ and $\beta$ two simple loops as in the construction of $\omega_{C;\alpha,\beta}$, if CLE events $\tou$ and $\t\tou$ are supported in $C_\beta$, then we have
\beq\label{CLEomega}
	P(\tou|\t\tou,\ev(\alpha,\beta))_C = \int P(\tou|\t\tou)_{C_\gamma} d\omega_{C;\alpha,\beta}(\gamma).
\eeq
\end{theorem}
\proof We need to use both the nesting axiom, for the possible loops that surround $\beta$ and intersect $\alpha$, and the conformal restriction axiom, for the loops that do not surround $\beta$ and intersect $\alpha$. In order to use the the nesting axiom, or the nesting property as expressed through re-randomisation in subsection \ref{ssectrandom}, we need to generalise it slightly. This is because we need to choose the loop $\gamma$ in which we re-randomise based on the fact that the configuration of loops inside it satisfies the conditions of a certain event $\toua$; the choice is not independent of the configuration inside the loop chosen. The event $\toua$ is the event that no loop surround $\beta$ and intersect $\alpha$. Indeed we need to choose the ``last loop'' that interesects $\alpha$ and surrounds $\beta$, before we apply the conformal restriction axiom. The measure for the configurations inside $\gamma$ is then a CLE measure restricted to $\toua$. More precisely, for a given configuration $x$ and a loop $\gamma\subset x$, recall the sequence $\sam_C'(\gamma,x)$ discussed in the nesting property in subsection \ref{ssectrandom}. We consider the choice map $\Phi$, mapping configurations in a subset of $\conf_C$ to loops in $C$, defined by the fact that the loop chosen is the one, if any, that intersects $\alpha$ and surrounds $\beta$, and such that there is no loop inside it that satisfies these two conditions. We note that the choice map $\Phi$ has the properties that for $x\in\conf_C$, if $\Phi(x)=\gamma$ is defined, then 1) $\gamma\in x$ (that is, we choose a loop in $x$), 2) $\inter_{C_{\gamma}}(x) \in \toua$ (that is, the loop we choose is such that the configuration of loops inside it satisfies the conditions of $\toua$), and 3) $\Phi(x')=\gamma$ for any $x'\in\conf_C$ such that $\gamma\in x'$, $\exter_{C_\gamma}(x) = \exter_{C_\gamma}(x')$ and $\inter_{C_{\gamma}}(x')\in\toua$ (that is, the choice does not depend on what is in the interior of the chosen loop, as long as it satisfies the conditions of $\toua$). Taking the subset $M$ of all integers $m$ for which $\Phi(x_m^C)=\gamma_m$ is defined, we have
\beq\label{gennestprop}
    P_{\sam_C}(\tou) = \lim_{N\to\infty} N^{-1}\sum_{m=1}^N
    \lt\{\ba{ll} \pres(x_m^C,\tou) & (m\not\in M) \\
    P_{\sam_C'(\gamma_m,x_m^C)}(\tou|\toua(\gamma_m)) & (m\in M). \ea\rt.
\eeq
Likewise, in the restricted case, we have
\beq\label{gennestproprest}
    P_{\sam_C}(\tou|\tou') = \lim_{N\to\infty} N^{-1}\sum_{m=\in I_{\tou''}^{(N)}}
    \lt\{\ba{ll} \pres(x_m^C,\tou) & (m\not\in M) \\
    P_{\sam_C'(\gamma_m,x_m^C)}(\tou|\tou',\toua(\gamma_m)) & (m\in M). \ea\rt.
\eeq
In these expressions, $\toua(\gamma_m)$ is the event that no loop intersect $\alpha$ and surround $\beta$ in the domain bounded by $\gamma_m$. That these expressions hold is proved as follows. Consider the nesting property. If a loop $\gamma$ in a configuration $x$ is chosen, then we know that the set of all configurations with ``similar'' exteriors to $\gamma$ (and with a loop ``similar'' to $\gamma$) will give rise to a CLE measure on the interior of $\gamma$. If in this set we only take the configurations such that the set of loops inside $\gamma$ satisfies $\toua$, and we take all of such configurations, then this will give rise to a CLE measure restricted to $\toua$ on the interior of $\gamma$. The definition of the map $\Phi$ above implies that we take only and all such configurations. Since $\sam_C'(\gamma,x)$ is a sequence such that all configurations have the loop $\gamma$, and giving rise, for the configurations of loops inside $\gamma$, to a CLE measure, the sequence where we restrict to $\toua$ for the configurations of loops inside $\gamma$, gives rise there to a CLE measure restricted to $\toua$. For the restricted case, the same arguments hold with the CLE measure restricted to $\tou'$.

Now we consider $\tou' = \ev(\alpha,\beta)\cap \t\tou$ and the fact that $\tou$ is supported inside $C_\beta$. Then, we have $P_{\sam_C'(\gamma_m,x_m^C)}(\tou|\tou',\toua(\gamma_m)) = P(\tou|\t\tou,\ev(\alpha,\beta),\toua)_{C_{\gamma_m}}$. We can now use conformal restriction for evaluating the latter probability, based on $C_{\gamma_m}\setminus C_\alpha$. Thanks to the restriction to the event $\toua$ and to the event $\ev(\alpha,\beta)$, there is always a component of the actual domain of restriction that contains $\beta$. Its boundary is the loop $\gamma$ of the construction above. Putting together nesting and conformal restriction, so that the re-randomisation procedure (its restricted version, in order to account to $\t\tou$) is performed only after these two steps, we obtain (\ref{CLEomega}).
\eproof

Then, from this, we obtain continuity of supported events:
\begin{theorem}\label{theocontevent}
An event $\tou$ is continuous at any simply connected domain $C$ that includes its support, and the continuity statement is independent of the event involved for a given support.
\end{theorem}
\proof From theorem \ref{theoCLEomega}, we can write $P(\tou|\ev(\alpha,\beta))_C = \int P(\tou)_{C_\gamma} d\omega_{C;\alpha,\beta}(\gamma)$, where we must choose $\beta$ to surround $\supp(\tou)$ (and $\alpha$ surrounds $\beta$). The function $C\mapsto P(\tou|\ev(\alpha,\beta))_C$ is continuous by theorem \ref{theocontexp}, since $P(\tou)_{C_\gamma}$ is bounded and positive, and the continuity statement does not depend on $\tou$ since the bound is always 1. From $P(\tou)_{C} = P(\tou|\ev(\alpha,\beta))_{C} P(\ev(\alpha,\beta))_{C} + P(\tou,\non\ev(\alpha,\beta))_{C}$, we have
\[
	0\leq P(\tou)_{C} - P(\tou|\ev(\alpha,\beta))_{C} P(\ev(\alpha,\beta))_{C} \leq P(\non\ev(\alpha,\beta))_{C}
\]
so that, by the continuity result just stated and by continuity of $\ev(\alpha,\beta)$ and $\non\ev(\alpha,\beta)$,
\[
	0\leq \lim_{n\to\infty} P(\tou)_{g_n(C)} - P(\tou|\ev(\alpha,\beta))_{C} P(\ev(\alpha,\beta))_{C} \leq P(\non\ev(\alpha,\beta))_{C}.
\]
Writing $P(\tou,\ev(\alpha,\beta))_C = P(\tou)_C - P(\tou,\non\ev(\alpha,\beta))_C$ and using $0\leq P(\tou,\non\ev(\alpha,\beta))_C\leq P(\non\ev(\alpha,\beta))_C$, we have in fact
\[
	- P(\non\ev(\alpha,\beta))_C\leq \lim_{n\to\infty} P(\tou)_{g_n(C)} - P(\tou)_C \leq P(\non\ev(\alpha,\beta))_{C}.
\]
We can take $\alpha=\alpha_m$ in a sequence of loops such that $\alpha_m \subset C_{\alpha_{m+1}}$, so that $\ev(\alpha_m,\beta)$ is an increasing sequence of events (and $\non\ev(\alpha_m,\beta)$ is a decreasing sequence). With $\lim_{m\to\infty} \alpha_m = \p C$, we have $\lim_{m\to\infty} P(\non\ev(\alpha_m,\beta))_C = 0$ because there are almost surely no loops that intersect $\p C$. Hence by choosing $\alpha$ appropriately we can make $\lim_{n\to\infty} P(\tou)_{g_n(C)}$ as near as we want to $P(\tou)_C$. Also, we see that the choice of $\alpha$ is independent of the event $\tou$.
\eproof

The following two theorems, similar to the above but slightly more technical, will be of use in proving crucial theorems of section \ref{sectdouble}. They mainly have to do with the properties of the continuity statements in the case of restricted probabilities, and in a similar case where a probability is affected by a factor. These subtleties turn out to be essential.

The first one deals with restricted probabilities, and will be crucial for the proof of theorem \ref{thcr2}, for instance.
\begin{theorem}\label{theores1}
The restricted probability $P(\tou|\tou',\ev(\alpha,\beta))_C$, where $\alpha$ surrounds $\beta$ which surrounds $\supp(\tou\cap\tou')$ and with $\tou'$ non-zero on its support, is continuous (as a function of $C$) at any simply connected domain $C$ that includes $\alpha$, and the continuity statement is independent of the events $\tou$ and $\tou'$.
\end{theorem}
\proof We use theorem \ref{theoCLEomega}, so that
\[
	P(\tou|\tou',\ev(\alpha,\beta))_C = \int P(\tou|\tou')_{C_\gamma} d\omega_{C;\alpha,\beta}(\gamma).
\]
The restricted probability $P(\tou|\tou')_{C_\gamma}$ is again bounded and positive, and the continuity statement does not depend on either $\tou$ or $\tou'$ since the bound is always 1.
\eproof

Note that the fact that the restricted probability is continuous is a simple consequence of theorem \ref{theocontevent}. However, the property of the continuity statement is not, so that the above theorem is non-trivial.

The second technical theorem is slightly more complicated to express, but is as simple to prove, and will be crucial for the proof of theorem \ref{corss}.
\begin{theorem}\label{theores2}
Consider a family of events $\tou_a$ for which there exists a common finite support, and a family of positive numbers $c_a$, with $a\in(0,1]$, such that $c_a P(\tou_a)_{\t{C}}$ has a finite supremum for $a\in(0,1]$, and for $\t{C}$ in a neighbourhood of a simply connected domain $C$ and containing the support. Consider also a simple loop $\alpha\in C$ which surrounds $\beta$ which surrounds the common support, where both $\alpha$ and $\beta$ are near enough to $\p C$. Then the function $\t{C}\mapsto c_a P(\tou_a|\ev(\alpha,\beta))_{\t{C}}$ is continuous at $\t{C} = C$, and the continuity statement only depends on $\tou_a$ and $c_a$ via the supremum of $c_a P(\tou_a)_{\t{C}}$.
\end{theorem}
\proof Using theorem \ref{theoCLEomega}, we have $c_a P(\tou_a|\ev(\alpha,\beta))_C = \int c_a P(\tou_a)_{C_\gamma} d\omega_{C;\alpha,\beta}(\gamma)$. Since $\beta$ is near enough to $\p C$, there is a supremum for $c_a P(\tou_a)_{C_\gamma}$ for $a\in(0,1]$ and for $\gamma$ supported by the measure. Hence, by theorem \ref{theocontexp}, the function $\t{C} \mapsto c_a P(\tou_a|\ev(\alpha,\beta))_C$ is continuous at $\t{C}=C$, and the continuity statement only depends on $\tou_a$ and $c_a$ through the supremum of $c_a P(\tou_a)_{\t{C}}$.
\eproof

\sect{CLE on $\C$}

The CLE measure was considered until now only on simply connected domains of $\C$. Its construction on $\C$ is a very non-trivial problem. It is not the purpose of this section (or of this work) to provide such a full construction, but rather to underline ideas as to how CLE probability functions on $\C$ may be obtained from those on simply connected domains. The main idea is that probability functions on $\C$ should be obtained from the very small loops around a point, scaled up so that they look ``finite''; or equivalently, from sending the boundary of the simply connected domain of definition to infinity. In order to do so, we will need an analysis of the small loops, then we will need to understand how global conformal invariance is recovered.

In this section we do not have complete proofs: we need to make two assumptions (or conjectures) about the CLE measure. The first assumption is about some general properties of the measure for the small loops: that they stay simple, and that they are distributed in some natural fashion. It is probable that a more precise look at the construction of CLE would provide a proof for this assumption, but here we just give some supporting arguments. The second assumption is likely to be harder to prove: it is an assumption of symmetry. As is explained below, it is a very natural assumption, easily seen to be satisfied in the underlying $O(n)$ model.

The construction of CLE probability functions on $\C$ will play an important r\^ole in the next section, where we consider probability functions on doubly connected domains. But also, it will play an essential r\^ole in the second part of this work, in the construction of the bulk stress-energy tensor.

\subsection{Local properties of CLE loops} \label{ssectlocal}

We first study what happens locally, in a neighbourhood of a point. We start with our main assumption, that tells us that looking at smaller and smaller loops, we find in the limit a measure on simple loops. The considerations of this assumption and of the theorems below have to do with loops surrounding the origin in CLE on $\uD$. In any configuration of $\sam_\uD$, consider the set of nested loops that surround the origin, $\{\gamma_n,\,n=1,2,3,\ldots\}$, with $\gamma_1$ the outermost. Consider the set of scales $\{\lambda^{(n)},\,n=1,2,3,\ldots\}$ such that the scaled loops $\lambda^{(n)} \gamma_n$ surround $\uD$ (that is, separate it from $\infty$) and touch $\p\uD$. Then:
\begin{assump} \label{asslimit} The set $\{\lambda^{(n)} \gamma_n,\,n=1,2,3,\ldots\}$ gives rise to a finite measure on simple loops, such that the conformal radius viewed from the origin has an induced measure that is absolutely continuous with respect to the Lebesgue measure on $[1,\infty)$. These measures are the same for any configuration of $\sam_\uD$.
\end{assump}
The meaning of the discrete set of loops giving rise to a measure is that for $n\to\infty$, we obtain a set of independent identically distributed (i.i.d.) random samples of the loop, governed by that measure (hence reproducing all averages), with an appropriate $\sigma$-algebra. The conformal radius viewed from a point $z$ inside a simple loop is defined as $\p g (0)$, where $g$ is the unique conformal transformation that maps the domain bounded by the loop and including $z$, to $\uD$, with $z\mapsto0$ and $\p g(0)>0$. It is greater than or equal to 1 because the loop surrounds $\uD$. The condition on its measure is essentially saying that it is a probability measure, and that the probability distribution is nice enough. As for the existence of the probability measure on the loops themselves, essentially, the process of starting with a domain $C\supseteq \uD$, constructing the CLE outer loop around the origin, and scaling it outside $\uD$ to get, through the nesting property, a new CLE domain $C$, is a Markov chain on the simple loops $\p C$. With appropriate properties of the corresponding Markov operator, the limit of a large Markov chain exists, and defines a measure on simple loops surrounding $\uD$ and touching $\p\uD$. Also, it does not matter which configuration we looked at, since the Markov operator is determined by the CLE measure only.

From assumption \ref{asslimit}, we may prove a useful theorem about how the loops around the origin are distributed.
\begin{theorem} \label{theoratio}
The set $\{\lambda^{(n)}\}$ gives rise to a finite measure for the ratio between two successive scales $\lambda^{(n)}/\lambda^{(n-1)}$ that is absolutely continuous with respect to the Lebesgue measure on $[1,\infty)$. This measure is the same for any configuration of $\sam_\uD$.
\end{theorem}
\proof  Let us consider the $n^{\rm th}$ loop $\gamma_n$. It can be parametrised by a conformal map $g_n$ from $\uD$ to the domain it bounds, fixing the origin. By the nesting property, this map can be written $g_n = g_{C_1} \circ \ldots \circ g_{C_n}$, where $g_C$ is a map from $\uD$ to $C$, with $\p C_n$ forming a sequence of independent samples of CLE outermost loops surrounding the origin, and with $g_C(0) =0$. We may take the normalisation $g_C(z) = \mu z + O(z^2)$ with $\mu>0$. In particular, the $\mu_i$ are i.i.d.\ random variables, and the measure on $\mu_i$ is finite and absolutely continuous with respect to the Lebesgue measure on $[0,1]$ (it is in fact known exactly, \cite{SSW06}). We have $g_n(z) = \mu_1 \cdots \mu_n z + O(z^2)$. By assumption \ref{asslimit}, we have that the measure obtained from the sequence $\lambda^{(n)} \mu_1\cdots \mu_n$ is finite and absolutely continuous with respect to the Lebesgue measure on $[1,\infty)$, and by construction we also have $\lambda^{(n)} > \lambda^{(n-1)}$. Taking the ratios $\lambda^{(n)} \mu_1\cdots \mu_n / (\lambda^{(n-1)} \mu_1\cdots \mu_{n-1}) = \mu_n\lambda^{(n)}/\lambda^{(n-1)}$, this completes the proof.
\eproof

In order to define a CLE probability function for events on $\C$, we construct a measure on simple loops derived from the measure of assumption \ref{asslimit}. There are two ways of constructing it: it occurs from any given CLE configuration in $\sam_\uD$, and it is induced by the CLE measure itself.

It will be convenient to have, for any closed set $B$ in some family, any positive number $\lambda$, and any CLE configuration $x$ on $\uD$, a choice of a loop $\gamma_\lambda^B\in x\cup \{\p\uD\}$. Consider a closed set $B$, not necessarily lying in $\uD$, with the condition that it be of non-zero extent (in this context, this means ${\rm max}_{z_1,z_2} d(z_1,z_2)>0$ where $d(z_1,z_2)$ is the Eucleudian distance on $\R^2$) and that it do not contain $\infty$. Consider the smallest closed disk $D_B$ centered at the origin that contains $B$. If there is at least one loop in $x$ that surrounds $\lambda D_B$, then the loop $\gamma_\lambda^B$ is the nearest to $\lambda D_B$. If there is no such loop, we set $\gamma_\lambda^B=\p\uD$. It is through the loop $\gamma_\lambda^B$ that we will construct a measure on simple loops.

First, from any given fixed configuration, we simply look at the loop $\gamma_\lambda^B$, scale down the set $B$ and scale up the loop in a way similar to assumption \ref{asslimit}, and put a Lebesgue measure on the logarithm of this scale.
\begin{theorem} \label{theolimitII} Consider any configuration of $\sam_\uD$, and the loop $\gamma_\lambda^B$ for a given set $B$. With the measure on $\lambda$ given by $d\lambda/\lambda$, and for any $\lambda'$ with $0<\lambda'<\infty$, the set $\{\lambda^{-1}\gamma_\lambda^B,\,\lambda\in(0,\lambda']\}$ gives rise to a $B$-dependent finite measure $\nu_B$ on simple loops $\gamma$ that separate $D_B$ from $\infty$ (that is, there is no continuous path from $\infty$ to any point of $D_B$ that does not intersect $\gamma$, and $\{\infty\} \cup D_B \subset \C\setminus \gamma$). This measure is the same for any $\lambda'$ and for any configuration of $\sam_\uD$.
\end{theorem}
\proof We may consider $\lambda$ small enough so that there is at least one such loop surrounding $\lambda D_B$ by propositions \ref{propoeverypoint} and the finiteness I property. Since the measure $d\lambda/\lambda$ has infinite weight on any open interval with endpoint at $\lambda=0$, this consideration does not affect the resulting measure on loops. As $\lambda$ is made smaller, there are more and more such loops by the nesting property. As we decrease $\lambda$, let us look at the nearest loop $\gamma_\lambda^B$ that surrounds $\lambda D_B$, and rescale it to $\lambda^{-1}\gamma_\lambda^B$ (the rescaled loop surrounds $D_B$). As we vary $\lambda$, the nearest loop stays the same, except when $\lambda^{-1}$ hits one of the values $R\lambda^{(n)},\,n=1,2,3,\ldots$ for $R$ the radius of $D_B$. Then we get a new nearest loop. By assumption \ref{asslimit}, the set of all new nearest loops for $\lambda\in(0,\lambda_{\rm max}^B)$ gives rise to a finite measure on the rescaled loops. By theorem \ref{theoratio}, the rescaling between different nearest loops is almost surely finite with respect to this measure. Hence, the rescaling covers almost surely a finite support of the measure $d\lambda/\lambda$. Then, by considering all $\lambda\in(0,\lambda_{\rm max}^B)$ with the measure $d\lambda/\lambda$, we have a finite measure on the rescaled loops, whose properties follow from assumption \ref{asslimit}.
\eproof

Second, we look at the measure on the scaled loops $\gamma_\lambda^B$ induced by the CLE measure, and show that it has a limit $\lambda\to0$, equal to the measure $\nu_B$. That is, we can look at a fixed small $\lambda$ and at all configurations, instead of a fixed configuration and all $\lambda$ small enough.
\begin{theorem} \label{theolimitIII} Consider the loop $\gamma_\lambda^B$ for a given set $B$. The CLE measure induces a measure on the scaled loop $\lambda^{-1} \gamma_\lambda^B$ that has a limit as $\lambda\to0$ given by $\nu_B$.
\end{theorem}
\proof Let us introduce the parameter $r=-\log\lambda$ and the random function of $n=1,2,3,\ldots$ defined by $t_n = \log(R\lambda^{(n)})$,  where $R$ is the radius of $D_B$. The nearest loop $\gamma_\lambda^B$ of the theorem is the $n^{\rm th}$ loop $\gamma_n$ where $n$ is the random variable defined by $t_n\leq r <t_{n+1}$. For any fixed $\lambda$, CLE induces a measure on the scaled loop $\lambda^{-1} \gamma_n$. Let us consider the random variable $s=r-t_n$. In the limit where $\lambda\to0$ the variable $n$ tends to infinity almost surely thanks to theorem \ref{theoratio}, so that the measure on $\lambda^{(n)}\gamma_n$ tends to the measure of assumption \ref{asslimit}. Since the scaled loop is $\lambda^{-1}\gamma_n = Re^{s} \lambda^{(n)}\gamma_n$, we need to understand the random variable $s$. Thanks to theorem \ref{theoratio}, the measure on the random variable $u_n = t_{n+1}-t_n$ has a limit $n\to\infty$ that is finite and absolutely continuous with respect to the Lebesgue measure; let us denote the corresponding random variable by $u$ (note that it may be correlated with the random loop $\lambda^{(n)}\gamma_n$). In order to reproduce the measure $\nu_B$, the variable $s$ must be determined as follows: for every instance of $u$, we must choose a $s$ in $[0,u)$ according to the (scaled) Lebesgue measure $ds/u$. This reproduces the measure $d\lambda/\lambda$ used in the construction of $\nu_B$ in theorem \ref{theolimitII}. By definition, we have $s\in[0,u)$. We may consider $t_n$, for $n$ large, essentially as the sum $\sum_{j=1}^{n-1} u_j$ of a very large number of i.i.d.\ random variables (independent from, and distributed like, $u$). Then we may rephrase the problem as follows. We add many i.i.d.\ random variables until the sum is larger than $r$. We denote the last random variable by $u$ and consider the difference between $r$ and the sum before the last random variable has been added, $s=r-t_n$. By ergodicity, for fixed $r$ and conditioned on $u$, the variable $s$ is uniformly distributed. This shows that it is the Lebesgue measure that results in the interval $s\in[0,u)$.
\eproof

This measure has the following covariance property:
\begin{corol}\label{corolinv}
The measure $\nu_B$ of theorems \ref{theolimitII}, \ref{theolimitIII} is covariant under scaling and rotation about the origin: $\nu_{a B} \circ a = \nu_B$ for $a\in\C,\,0<|a|<\infty$.
\end{corol}
\proof Scaling covariance is a direct consequence of the invariance of $d\lambda/\lambda$ under scaling and of the independence from $\lambda'$ in theorem \ref{theolimitII}, or simply of the fact that the limit $\lambda\to0$ exists in theorem \ref{theolimitIII}. Rotation covariance is a consequence of rotation invariance of the CLE measure on $\uD$, and of the fact that rotations commute with the scale transformations used in the definition of the measure $\nu_B$.
\eproof

In the following we will often need to take averages of functions $f(\t{C})$ of a domain $\t{C}$ containing the origin and bounded by the random simple loop controlled by the measure $\nu_B$. We will denote the average as follows:
\beq
	\mathbb{E}_{\p\t{C} \,:\,\nu_B} [f(\t{C})] \equiv \bra f(\t{C})\ket_B
\eeq
(that is, we take $\t{C}$ implicitly as the random domain).

\subsection{Probability function for CLE on $\C$} \label{subsectCLEC}

The local properties of loops bring us some way towards defining the CLE probability function on $\C$. Essentially, the measure $\nu_B$ above can be used for this purpose. The basis for this is the next theorem.
\begin{theorem}\label{thlimC}
Consider $\tou$ and $\tou'$ some events with $\supp(\tou) \subset C,\,\supp(\tou')\subset C$ where $C$ is a simply connected domain, and $\tou'$ non-zero on its support. Consider $z\in C$, and $z'\not\in C$, such that $\tou'$ is supported away from $z$. Then, the following limit exists and gives:
\beq\label{tlCdl}
	\lim_{\lambda\to0} P(\lambda_{z,z'}\tou|\tou')_C
	=\bra P(h\tou)_{\t{C}} \ket_{h(\supp(\tou))}.
\eeq
Here, $h$ is any global conformal map that maps $z'$ to $\infty$. In particular, for any $C$, $\tou'$, $z$ and $z'$, the limit (\ref{tlCdl}) may only depend on $z'$.
\end{theorem}
\proof We will use theorem \ref{theolimitIII}. Let us first consider the case where $\tou'$ is the trivial event. Let us use a conformal transformation $g$ to map the problem onto $\uD$, with $z\mapsto0$. For simplicity, we will first assume $z\neq\infty$. Then, we may choose $\p g(z) >0$ (note that for a transformation conformal at $z$, we have $\p g(z)\neq0$). We have, using (\ref{relation1}),
\beqa
	P(\lambda_{z,z'} \tou)_C &=& P(g\circ \lambda_{z,z'} \tou)_\uD \n
	&=& \mathbb{E}_{\gamma_\lambda^B} \lt[P(g\circ\lambda_{z,z'} \tou)_{\uD_{\gamma_\lambda^B}}\rt] \n
	&=& \mathbb{E}_{\gamma_\lambda^B} \lt[P(\lambda^{-1} \circ g\circ\lambda_{z,z'} \tou)_{\lambda^{-1}\uD_{\gamma_\lambda^B}}\rt]
\eeqa
where $\gamma_\lambda^B$, determined by the CLE measure, is as defined in subsection \ref{ssectlocal}, with $B= \lambda^{-1}\circ g\circ\lambda_{z,z'}(\supp(\tou))$.

Using (\ref{genscale}), we have that $\lambda^{-1} \circ g \circ \lambda_{z,z'}$ is uniformly convergent on $\supp(\tou)$ as $\lambda\to0$. Indeed, for any $x\in\supp(\tou)$,
\beqa
	\lambda^{-1} \circ g\circ \lambda_{z,z'}(x) &=& \lambda^{-1} g\lt(z + \lambda (z-z')\frc{x-z}{x-z'} + O(\lambda^2)\rt) \n
		&=& \p g(z) (z-z') \frc{x-z}{x-z'} + O(\lambda).
\eeqa
To leading order, this is a global conformal map, hence the support transforms accordingly by corollary \ref{coroltranssupp}. Hence by continuity, theorem \ref{theocontevent}, $\lim_{\lambda\to0} P(\lambda^{-1} g\circ\lambda_{z,z'} \tou)_{\t{C}}$ exists for any simply connected domain $\t{C}$ that contains the transformed support and excludes $\infty$, and gives $P(\p g(z) (z-z') h\tou)_{\t{C}}$, where
\beq
	h(x) = \frc{x-z}{x-z'}.
\eeq
This implies that $P(\lambda^{-1}\circ g\circ\lambda_{z,z'} \tou)_{\t{C}}$ converges almost everywhere as $\lambda\to0$, with respect to the measure on $\t{C}=\lambda^{-1}\uD_{\gamma_\lambda^B}$ taken at any fixed $\lambda$, and since it is bounded by 1, Lebesgue's bounded convergence theorem \cite{H50} implies that the average on $\t{C}$ converges to the average of its limit.

We now want to take the limit $\lambda\to0$ on the measure on $\t{C}$. We may choose $B$ to be $\lambda$-independent and big enough so that it contains $\lambda^{-1}\circ g\circ\lambda_{z,z'}(\supp(\tou))$ for all $\lambda$ small enough, and apply theorem \ref{theolimitIII} to obtain the measure $\nu_B$ in the limit $\lambda\to0$. Hence, we have
\beq
	\lim_{\lambda\to0} P(\lambda_{z,z'}\tou)_C = \bra P(\p g(z) (z-z') h\tou)_{\t{C}} \ket_{\p g(z) (z-z') h(\supp(\tou))}.
\eeq
By corollary \ref{corolinv}, we have $\bra f(a\t{C})\ket_{A} = \bra f(\t{C})\ket_{aA}$ for $a\in\C,\,0<|a|<\infty$ (for any allowed $A$). Hence, we find
\beq\label{eqfppr}
	\lim_{\lambda\to0} P(\lambda_{z,z'}\tou)_C = \bra P(h\tou)_{\t{C}} \ket_{h(\supp(\tou))}.
\eeq
Hence, the limit is independent of $C$.

Let us now consider the case where $\tou'$ is non-trivial. Consider the event $\Gamma_{\lambda'}$ that at least one loop $\gamma$ surrounds $\lambda_{z,z'}'(\supp(\tou))$, and separates it from $\supp(\tou')$, for some $\lambda'>0$ small enough. We may use (\ref{relation4}), and we find
\beqa
	P(\lambda_{z,z'} \tou,\tou')_C &=& P(\lambda_{z,z'} \tou,\tou',\Gamma_{\lambda'})_C +
		P(\lambda_{z,z'} \tou,\tou',\non \Gamma_{\lambda'})_C \\
	&=& \lim_{N\to\infty} N^{-1}\sum_m P(\lambda_{z,z'} \tou)_{C_{\gamma_m}} \pres(x_m^C,\tou') P(\Gamma_{\lambda'}) +
		P(\lambda_{z,z'} \tou,\tou',\non \Gamma_{\lambda'})_C \no
\eeqa
where the sum over $m$ takes the first $N$ configurations of $\sam_C$ that satisfy the conditions of the event $\Gamma_{\lambda'}$, and $\gamma_m$ is, in the configuration $m$, the biggest loop that fulfills the conditions of $\Gamma_{\lambda'}$. We may take the limit $\lambda\to0$, using the fact that it is independent of $C_{\gamma_m}$:
\beq
	\lim_{\lambda\to0} \lt(P(\lambda_{z,z'} \tou,\tou')_C - P(\lambda_{z,z'} \tou,\tou',\non \Gamma_{\lambda'})_C\rt)
	= \bra P(h\tou)_{\t{C}}\ket_{h(\supp(\tou))} P(\tou',\Gamma_{\lambda'})_C.
\eeq
On the other hand,
\beq
	0 \leq P(\lambda_{z,z'} \tou,\tou',\non \Gamma_{\lambda'})_C \leq P(\non\Gamma_{\lambda'})_C
\eeq
so that we have the bounds
\beq
	0 \leq \lim_{\lambda\to0} P(\lambda_{z,z'} \tou,\tou')_C - \bra P(h\tou)_{\t{C}}\ket_{h(\supp(\tou))} P(\tou',\Gamma_{\lambda'})_C \leq
	P(\non\Gamma_{\lambda'})_C.
\eeq
But by corollary \ref{corolsurr}, we have $\lim_{\lambda'\to0} P(\Gamma_{\lambda'}) = 1$, so that
\beq\label{restlC}
	\lim_{\lambda\to0} P(\lambda_{z,z'} \tou|\tou')_C = \bra P(h\tou)_{\t{C}}\ket_{h(\supp(\tou))}.
\eeq
Hence this is independent of $\tou'$.

The right-hand side of (\ref{restlC}) is independent of the event $\tou'$ and of $C$, as long as $\supp(\tou)\cup \supp(\tou') \in C$, $z\in C$ and $z'\not\in C$, and with the condition $z\neq \infty$. Note also that our choice of support is arbitrary, since the left-hand side is independent of it, as long as it is a good support as required by the event $\tou$. We may denote the quantity (\ref{restlC}) by
\[
	Q(z,z',\tou).
\]
Conjugating $\lambda_{z,z'}$ by a global conformal transformation $G$, we obtain $G\circ\lambda_{z,z'}\circ G^{-1} =\lambda_{G(z),G(z')}$, and this gives us the covariance property
\beq\label{0tlC}
	Q(G(z),G(z'),G\tou) = Q(z,z',\tou)\quad \mbox{($G$ a global conformal transformation).}
\eeq
In the probability function on the left-hand side of (\ref{restlC}), let us consider $z=0,\,z'=\infty$ and $\tou= G^{-1}\tou''$ for $G(x) = b/(cx+\delta),\,b,c,\delta\in\C\setminus\{\infty\}$ (with $b\neq0,c\neq0$). Conjugating, we find
\beq\label{1tlC}
	Q(0,\infty,G^{-1}\tou'') = Q(b/\delta,0,\tou'').
\eeq
On the other hand, we may apply the $\lambda$-dependent conformal transformation $G_\lambda(x) = b/(cx+\lambda \delta)$ on both the domain of definition $C$ and the event inside the probability function, on the left-hand side of (\ref{restlC}). The limit $\lambda\to0$ of $G_\lambda(C)$ gives a non-empty domain, and $G_\lambda\circ \lambda_{0,\infty} = \lambda_{\infty,0}\circ G$. Hence the limit exists and is independent of $C$, and we find
\beq\label{2tlC}
	Q(0,\infty,G^{-1}\tou'') = Q(\infty,0,\tou'').
\eeq
Equality between (\ref{1tlC}) and (\ref{2tlC}), along with the covariance property (\ref{0tlC}), means that $Q(z,z',\tou)$ is independent of $z$ for ;any $z'$, and that we can relax the condition $z\neq\infty$. Hence, we may take for $z$ any point in $C$, and (\ref{restlC}) gives (\ref{tlCdl}).
\eproof

From theorem \ref{thlimC}, we find invariance in the limit under any global conformal transformation that fixes $z'$. We would expect that the dependence upon the variable $z'$ should in fact disappear as well, from which invariance under general global conformal transformation would follow, as is required of a conformally invariant probability function on $\C$. It seems, however, that this is impossible to prove solely using the arguments above, about conformal invariance, the nesting property, and existence of limit measures. This is a crucial fact, and perhaps relates to the essential r\^ole of conformal restriction in CLE for obtaining the conformal field theory structure, in particular its relation to some concept of locality, at the basis of the properties of the stress-energy tensor. Note that there was no direct need for conformal restriction in the proof of the theorem above, and it may be possible to construct measures that do not have conformal restriction, yet that possess all appropriate properties in order for that theorem to hold.

In order to obtain global conformal invariance, we make the following non-trivial symmetry assumption:
\begin{assump}\label{assumpsym}
Consider an event $\tou$ with $i\not\in\supp(\tou),\,-i\not\in\supp(\tou)$, and $\supp(\tou)$ included inside a domain. Then, with $h(z) = (z-i)/(z+i)$, the averaged probability $\bra P(h\tou)_{\t{C}}\ket_{h(\supp(\tou))}$ is mirror symmetric with respect to the real line:
\beq\label{symeq}
	\bra P(h\tou)_{\t{C}}\ket_{h(\supp(\tou)\cup \{i\})} = \bra P(h^*\tou)_{\t{C}}\ket_{h^*(\supp(\tou)\cup \{-i\})}
\eeq
where $h^*(z) = (z+i)/(z-i)$.
\end{assump}
On the left-hand side of (\ref{symeq}) the random domain $h^{-1}(\t{C})$ contains $\supp(\tou)\cup \{i\}$, and keep $-i$ in the exterior. The assertion is that this gives the same result as using a random domain containing $\supp(\tou)\cup\{-i\}$ and keeping $i$ in the exterior. The idea is that, for instance, the left-hand side is obtained, according to theorem \ref{thlimC}, by considering the probability of $\tou$ on a domain of definition containing $\supp(\tou)\cup\{i\}$ but not $-i$, and making this domain ``bigger and bigger'', that is, making its exterior around $-i$ smaller and smaller. When this domain is very big, the measure on the loops in the region of $\supp(\tou)$ and in any neighbourhood of $i$ is independent from the shape of the domain, as is suggested by theorem \ref{thlimC}. Then, we may, without changing the result, put a measure on the space of domains of definition $C$ chosen so that $\p C$ reproduces the mirror image of the measure on some small loop around $i$. But since the exterior of a loop is like the boundary of a domain -- essentially the content of the conformal restriction axiom --, then we may as well consider the small loop around $i$ as a domain boundary, and $\p C$ as a random loop. This gives rise to the right-hand side.

It is essential, in this argument supporting the assumption, that the event $\tou$ do not make explicit reference to the boundary of the domain. This was one of the reasons for defining events as subsets of $\conf_\C$, and only {\em a posteriori} imposing that $\tou_C$ be in $\sgm_C$ for simply connected domains $C$ containing the support.

It is likely to be very hard to prove this assumption by ``elementary'' means from the axioms of CLE, or even from the explicit constructions of the CLE measure in \cite{W05a,ShW07b}. However, this mirror symmetry is a very natural and expected property. For instance, it is easy to see that if indeed $\p C$ and the loop around $i$, in the discussion above, can be made to have mirror symmetric measures, then the symmetry holds for any {\em finite} measure on configurations of loops on a lattice with the factorisation property $\mu(\{\gamma_i\}_{i=1}^N) = \prod_{i=0}^N \mu(\gamma_i)$. The $O(n)$ model that is conjectured to give rise to the CLE measure in the continuum limit has this factorisation property, hence any proof of this conjecture should provide, through a sligthly stronger version of theorem \ref{thlimC}, a proof of the assumption above.

We may now prove independence from $z'$ in the set-up of theorem \ref{thlimC}:
\begin{theorem}\label{theoindep}
In the set-up of theorem \ref{thlimC}, the limit $\blim_{\lambda\to0} P(\lambda_{z,z'}\tou|\tou')_{C}$ is the same for any $z,\,z',\,C$ and $\tou'$ satisfying the conditions of the theorem.
\end{theorem}
\proof We show that with $z\in C$, $z'\in \C\setminus\overline{C}$ and $z,z' \not\in\supp(\tou)$, we have
\beq\label{eqthindep}
	\blim_{\lambda\to0} P(\tou)_{\lambda_{z',z}C} = \blim_{\lambda\to0} P(\tou)_{\lambda_{z,z'}(\C\setminus \overline{C})}.
\eeq
By covariance under global conformal transformations we can always choose $z=i,z'=-i$. Then, theorem \ref{thlimC}, corollary \ref{corolaugmsupp} and assumption \ref{assumpsym} proves (\ref{eqthindep}). Finally, since the left-hand side is independent of $z$, and the right-hand side is independent of $z'$, both are independent of $z$ and $z'$, which completes the proof. \eproof

The limit involved in theorem \ref{thlimC} is essentially what we define as the CLE probability function on $\C$:
\begin{defi}\label{defCplane}
The probability of an event $\tou\in\ev(\conf_\C)$ on $\C$, where $\tou$ possesses a support lying in some domain, is defined by
\beq\label{formula}
    P(\tou)_{\C} \equiv \bra P(h\tou)_{\t{C}} \ket_{h(\supp(\tou))}.
\eeq
Here, $\bra\cdot\ket_B$ is an average over $\p \t{C}$ according to the measure $\nu_B$ of theorems \ref{theolimitII}, \ref{theolimitIII}, and $h$ is any global conformal map that maps some point outside $\supp(\tou)$ to $\infty$.
\end{defi}
This definition gives indeed a probability function, and in particular the result is independent of $h$ thanks to theorems \ref{theoindep} and \ref{thlimC}. We note that if $\tou$ is non-zero on its support, where its support is contained inside some domain, then its probability on $\C$ is non-zero as well. We will say that $\tou$ is non-zero on $\C$.

A subtlety arises when considering restricted probabilities. The restricted probability associated to this probability function is naturally given by
\beq\label{firstrest}
	P(\tou|\tou')_\C = \frc{P(\tou,\tou')_\C}{P(\tou')_\C}
\eeq
for $\tou,\tou'$ events satisfying the conditions of the definition, and $\tou'$ non-zero on $\C$. The following theorem gives a different expression for the same object, based on restricted re-randomisation procedures:
\begin{theorem}
The restricted probability function on $\C$ of an event $\tou$ with respect to an event $\tou'$, both satisfying the conditions of definition \ref{defCplane} and with $\tou'$ non-zero on $\C$, is given by
\beq\label{formularest}
	P(\tou|\tou')_\C = \bra P(h\tou|h\tou')_{\t{C}} \ket_{h(\supp(\tou)\cup\supp(\tou'))}.
\eeq
\end{theorem}
\proof Starting with the expression of $P(\tou,\tou')_\C$ and $P(\tou')_\C$ in terms of limit through theorem \ref{thlimC}, we can write $P(\tou|\tou')_\C$ (\ref{firstrest}) as the limit of a restricted probability on a simply connected domain,
\[
	\lim_{\lambda\to0} P(\lambda_{z,z'}\tou|\lambda_{z,z'}\tou')_C.
\]
Then, we simply have to re-trace the first part of the proof of theorem \ref{thlimC}, up to (\ref{eqfppr}), but using (\ref{relation4rest}) (specialised to $\tou'$ the trivial event) instead of (\ref{relation1}), and using corollary \ref{corolaugmsupp}.
\eproof

We may now prove global conformal invariance for probabilities on $\C$:
\begin{theorem}\label{thglobinv}
Probabilities on $\C$ are invariant under global conformal transformations, i.e. conformal transformations $G:\C\to\C$ with $z\mapsto (az+b)/(cz+d)$, $ad-bc=1$. That is,
\beq
    P(G(\tou))_\C = P(\tou)_\C.
\eeq
\end{theorem}
\proof For a conformal transformation $G$ that preserves $\C$, we have that $G(\lambda_{z,z'} C) \subset \C$ for all $\lambda>0$, and also $G\circ \lambda_{z,z'} (C) = \lambda_{G(z),G(z')}\circ G(C)$. Hence, using theorem \ref{theoindep}, we have
\beqa
	P(\tou)_{\C} &=& \blim_{\lambda\to0} P(\lambda_{z,z'}\tou)_{ C} \n
		&=& \blim_{\lambda\to0} P(G\circ \lambda_{z,z'}\tou)_{G( C)} \n
		&=& \blim_{\lambda\to0} P(\lambda_{G(z),G(z')}\circ G\tou)_{ G(C))} \n
		&=& P(G\tou)_\C.\no
\eeqa
\eproof

Finally, we have a theorem of factorisation:
\begin{theorem}\label{thfact}
Consider $\tou$ and $\tou'$ two events with $\supp(\tou) \subset C,\,\supp(\tou')\subset C$ for $C$ a simply connected domain. Consider two points $z\in C$ and $z'\not\in C$, such that $\tou'$ is supported away from $z$. Then,
\beq
	\blim_{\lambda\to0} P(\lambda_{z,z'}\tou,\tou')_C = P(\tou)_\C P(\tou')_C.
\eeq
This formula also holds for $z,z'\in\C$, $z\not\in\supp(\tou')$, $z'\not\in \{z\}\cup \supp(\tou)$ with $C=\C$.
\end{theorem}
\proof In the first case, this is a direct consequence of theorem \ref{thlimC} and definition \ref{defCplane}. In the second case, let us consider $z'=\infty$ without loss of generality by global conformal invariance, theorem \ref{thglobinv}. Then, we can use definition \ref{defCplane} with $h={\rm id}$, and we have
\beq
	\blim_{\lambda\to0}\bra P(\lambda_{z,\infty}\tou,\tou')_{\t{C}} \ket_{\supp(\tou)\cup\supp(\tou')}.
\eeq
The limit can be taken inside the average, since almost surely $\t{C}$ is in agreement with the conditions of theorem \ref{thlimC}, and using the fact that $P(\lambda_{z,\infty}\tou,\tou')_{\t{C}}$ is bounded by 1 so that Lebesgue's bounded convergence theorem applies. This gives
\beq
	\bra P(\tou)_\C P(\tou')_{\t{C}} \ket_{\supp(\tou)\cup\supp(\tou')} = P(\tou)_\C P(\tou')_\C
\eeq
which completes the proof in the second case. \eproof

\section{CLE on doubly connected domains}\label{sectdouble}

Through the use of the events $\ev(\alpha,\beta)$ defined in subsection \ref{subsectevents} and the use of the CLE probability function on $\C$ considered in the previous section, we are now in a position to provide some of the main ideas for defining a CLE probability function on annular domains (i.e.\ doubly connected domains). There is one additional assumption made in this section, but the theorems rely on the assumptions of the previous section only through the properties of the probability function on $\C$ derived there. Hence, an independent derivation of the properties on $\C$, where for instance the assumptions of the previous section are not proven, would be enough for this section.

In the perspective of constructing a probability function on annular domains, it will be convenient to have a different characterisation of the event $\ev(\alpha,\beta)$, making the annular region clearer (see figure \ref{figtheloop} in subsection \ref{subsectsupport}). For a simply connected domain $A$, a positive real number $\varep$ and a continuous function $u: \p A\to \C$, let us construct a {\em partner} $B$ of $A$ as a simply connected domain that is completely included inside $A$, with $\p B$ disjoint from $\p A$, and that is bounded by $\p B = \lt(\varep u + {\rm id}\rt)(\p A)$. Not all $u$ and $\varep>0$ give such a partner, but let us consider all $u$ such that for any $\varep>0$ small enough, $B\subset A,\,\p B\cap \p A = \emptyset$. We will denote by $\ev(A,\varep,u)$ the event that all loops that intersect $\cl{B}$ are completely included inside $A$. Using our previous notation, we have $\ev(A,\varep,u) = \ev(\p A,\p B)$. For convenience, we will require $\infty\not\in\p A$. In other words, $\ev(A,\varep,u)$ is the event asking that no loop cuts transversally the ``fattened'' boundary of $A$, of width of the order of $\varep$; this fattened boundary, for $\varep>0$ small enough, does not contain $\infty$. The restriction that $\infty\not\in\p A$ is only for technical simplifications in the discussion below. Note that the probability of such events vanishes as $\varep\to0$ (as the fattening goes to zero), because there will be almost surely at least one small loop that breaks the condition of the event. In fact, it is easy and useful to consider more generally functions $u$ that depend smoothly on $\varep$. We will impose the condition that as $\varep\to0$, the function $u$ tends uniformly, on $\p A$, to a continuous function $u_0$ that, if put in place of $u$, gives a partner for all $\varep$ small enough.

The first theorem shows how $\ev(A,\varep,u)$ may be used to make $A$ a new domain of definition for the CLE probability.
\begin{theorem}\label{thcr1}
Consider $C$ a simply connected domain or $C=\C$. With $A\subset C$ a simply connected domain and $\tou$ supported inside $A$, we have
\beq\label{eqcr1}
	\lim_{\varep\to0} P(\tou|\ev(A,\varep,u))_C = P(\tou)_{A}.
\eeq
With $\C\bs \overline{A}\subset C$ a simply connected domain and $\tou$ supported inside $\C\bs\overline{A}$, we have similarly
\beq
	\lim_{\varep\to0} P(\tou|\ev(A,\varep,u))_C = P(\tou)_{\C\bs\overline{A}}.
\eeq
\end{theorem}
\proof Let us consider any configuration $x_j^C$ in $\sam_C$ that satisfies the conditions of $\ev(A,\varep,u)$. Consider the first part of the theorem. For any fixed $\varep>0$ small enough, we are looking at the restriction that no loop intersects both $\C\bs A$, and the closure of the partner of $A$, $\overline{B_\varep}\subset A$. Let us consider nesting and conformal restriction based on $A$, in order to construct the loop $\gamma$ as in figure \ref{figtheloop}. It bounds a domain $C_\gamma$ that includes $B_\varep$. For $\varep$ small enough (that does not depend on the configuration we are looking at), $\supp(\tou)\subset B_\varep$, and the contribution can be written, from theorem \ref{theoCLEomega}, $P(\tou)_{C_\gamma} = P(\tou)_{g_\gamma(A)}$, with $g_\gamma:A\to C_\gamma$. As $\varep\to0$, we have $C_\gamma\to A$, and $g_\gamma$ tends to the identity. Hence by continuity (theorem \ref{theocontevent}), the limit exists and is given by $P(\tou)_{A}$ on every configuration. Since this limit obviously has the finite average $P(\tou)_A$ when looking at all configurations, and since $P(\tou)_{C_\gamma}$ is always bounded by 1, the limit exists on the initial probability function by Lebesgue bounded convergence theorem and gives (\ref{eqcr1}). This shows the first part of the theorem with $C\subset\C$ a simply connected domain. With $C=\C$, we need to use (\ref{formularest}) (without loss of generality, we can always assume $\infty$ to be away from $A$, so we can take $h=\id$). Then, the same arguments hold for every sample of the random domain $\t{C}$, and since the result is independent of $\t{C}$, the average over $\t{C}$ of the limit also exists and gives the same result. Finally, for the second part of the theorem, with $\C\bs \overline{A}\subset C$, the proof is similar; the only difference is that conformal restriction is based on the intersection between $C$ and the complement of the partner of $A$.
\eproof

The fact that the limit in the theorem above exists and gives the result stated is natural: we forbid loops from crossing the ``fattened'' boundary of $A$, and as this fattening goes to 0, we are in effect separating $A$ from its outside in $C$. Since the event $\tou$ is supported on $A$, we are left with a probability function for CLE on $A$. Note that for this purpose, we could have used, instead, the event that at least one loop is present that ``goes around'' inside the fattened boundary of $A$. The same result would have been obtained using the nesting property of CLE instead of theorem \ref{theoCLEomega}. However, as we said before, the particularities of the event $\ev(A,\varep,u)$ are used in other situations. Also, as will be discussed in the second part of this work, it is more natural, from the viewpoint of the stress-energy tensor, to use the event $\ev(A,\varep,u)$, because it generalised more naturally to the construction of the boundary stress-energy tensor.

In the theorem above, in order to be able to express the result of the limit as a CLE probability on a simply connected domain, we needed the condition that $\tou$ be supported on $A$, but certainly the existence of the limit should only depend on local properties of the small loops near to $\p A$. That is, the limit should exist also if $\tou$ is supported in the annular region $C\bs \overline{A}$. Moreover, like in the case where $\tou$ is supported on $A$, this limit should not depend on the precise fattening function $u$ chosen. The result could be naturally interpreted as a probability function on the annular region $C\setminus \overline{A}$. In this sense, then, the result should be the same as that of the limit $\varep\to0$ of a CLE probability function on $\C\setminus \overline{A}$, restricted to the event $\ev(C,\varep,u)$. These statements are essentially the content of the next theorem.

However, in order to establish it, we need one assumption about the CLE measure. It is a somewhat weak assumption, but it would require an independent proof. We know that in CLE, the measure on the set of configurations where one or many loops touch the boundary of the domain of definition is zero. This is clear from the fact that the CLE measure is {\em a priori} defined on a set of configurations with the property that no loop touches the boundary (so the set of configurations where one or many loops touch the boundary is the empty set). However, we wish to be able to take limits, so that we must consider the completion of this space, and of the CLE $\sigma$-algebra on it. Then, the set of configurations where one or many loops touch the boundary of the domain of definition is non-zero, but has zero measure (by completion). Now, let us consider the limit $\lim_{\varep\to0} \ev(A,\varep,u)$ as a limit on sets. It exists, because $\ev(A,\varep,u)$ are decreasing sets, so that it equals $\cap_{\varep>0} \ev(A,\varep,u)$. We may obtain a $\sigma$-algebra of subsets of $\cap_{\varep>0} \ev(A,\varep,u)$ by considering the limit $\lim_{\varep\to0} \tou\cap \ev(A,\varep,u)$ for $\tou$ in the $\sigma$-algebra of CLE on some domain. This limit also exists. There is also a natural measure induced on this $\sigma$-algebra by the CLE measure on $C$ with $\p A\subset C$. One can expect that this should be a product of two measures and should include, as a factor, the measure for CLE on annular domains. The assumption is that in the induced measure, the loops still almost surely do not touch the boundary of the initial domain of definition $\p C$. This is very natural, since the event $\ev(A,\varep,u)$ only imposes conditions to loops that go near to $\p A$, which is far from the boundary of the domain of definition; the event $\ev(A,\varep,u)$, then, should not select only the configurations where one or many loops are imposed to go near to the boundary of the domain of definition. Likewise, we can also construct an induced measure from the limits $\lim_{\varep\to0} \lim_{\varep'\to0}  \ev(A,\varep,u)\ev(B,\varep',u')$ with $\p B\subset C$ and $\p A\subset C$ and $\p B,\,\p A$ disjoint; we expect that in this induced measure, the loops almost surely do not touch $\p C$. Other orders of limits are also possible, with the same statement about the induced measure (although the induced measure may be different).
\begin{assump}\label{assborder}
In the induced measure, from the CLE measure on $C$, on $\cap_{\varep>0} \ev(A,\varep,u)$, and in the induced measures on $\cap_{\varep>0,\,\varep'>0} \ev(A,\varep,u)\cap \ev(B,\varep',u')$ obtained by any possible orders of the limits, with $\p A\subset C$, $\p B\subset C$ and $\p A,\,\p B$ disjoint, the loops almost surely do not touch the boundary $\p C$ of the domain of definition.
\end{assump}

It is interesting to note that, from this, the following theorem shows that the loops also almost surely do not touch $\p A$ -- the other component of the boundary of the annular domain.
\begin{theorem}\label{thcr2}
Consider $A,\,C$ simply connected domains with $\cl{A}\subset C$. Consider $\tou$ supported inside $C\setminus \cl{A}$. We have
\beq\label{eqthcr2}
    \lim_{\varep\to0} P(\tou|\ev(A,\varep,u))_C = \lim_{\varep\to0} P(\tou|\ev(C,\varep,u'))_{\C\setminus \cl{A}}
\eeq
for any $u$ and $u'$ (appropriate for the events $\ev$ involved). In particular, both sides exist (in the ordinary sense of limits) and are independent of $u$ and $u'$.
\end{theorem}
\proof As a first step, let us consider instead the proof of a slightly different relation. We consider a simply connected domain $B$ ``in-between'' the domains $A$ and $C$, that is, with $A\subset B\subset C$ and $\p A,\,\p B,\,\p C$ not intersecting each other. Then, we first wish to prove
\beq\label{eqpr2}
    \lim_{\varep\to0} P(\tou|\ev(A,\varep,u),\ev(B,s,v))_C = \lim_{\varep\to0} P(\tou|\ev(B,s,v),\ev(C,\varep,u'))_{\C\setminus \cl{A}}
\eeq
where $s$ and $v$ are kept fixed, and in such a way that the partner of $B$ includes $\cl{A}$. Using theorem \ref{thcr1}, we write the left-hand side as
\[
	\lim_{\varep\to0} \lim_{\varep'\to0} P(\tou|\ev(A,\varep,u),\ev(B,s,v),\ev(C,\varep',u'))_\C
\]
and the right-hand side as
\[
	\lim_{\varep'\to0} \lim_{\varep\to0} P(\tou|\ev(A,\varep,u),\ev(B,s,v),\ev(C,\varep',u'))_\C.
\]
In these expressions, the first limit taken always exists. Hence let us consider
\[
	P(\tou|\ev(A,\varep,u),\ev(B,s,v),\ev(C,\varep',u'))_\C.
\]
We take the formulation of the restricted probability on $\C$ given by (\ref{formularest}), with the choice $C\subset\t{C}$. Let us consider any configuration $x_j^{\t{C}}$ in $\sam_{\t{C}}$ that satisfies the restriction of the probability. We construct the loop $\gamma$ in the fattened boundary of $C$ as in figure \ref{figtheloop}, and the domain $C_\gamma$. For $\varep'$ small enough (that does not depend on the configuration), the partner of $C$ contains $B$, and we get $P(\tou|\ev(A,\varep,u),\ev(B,s,v))_{C_\gamma}$. By theorem \ref{theores1}, this is continuous as a function of $C_\gamma$, and the continuity statement does not depend on $\tou$ or $\ev(A,\varep,u)$. In particular, it does not depend on $\varep$. The use of this theorem is why we needed to insert $\ev(B,s,v)$. Hence:
\beqa
	\lefteqn{\forall \delta>0:\; \exists \varep_0>0 \;|\; \forall \varep>0,\,\forall \varep'<\varep_0:}  && \n &&
	-\delta<P(\tou|\ev(A,\varep,u),\ev(B,s,v))_{C_\gamma}-P(\tou|\ev(A,\varep,u),\ev(B,s,v))_C<\delta \no
\eeqa
(recall that $C_\gamma$ tends to $C$ when the fatness $\varep'$ tends to 0). The value of $\varep_0$ may depend on the configuration, which changes the family of domains $C_\gamma$ for $\varep'>0$, but it may be chosen so as to be valid for all families in all configurations, because the distance between $\p C_\gamma$ and $\p C$ has a supremum. Hence, we can average over $C_\gamma$ and $\t{C}$, in which case
\[
	P(\tou|\ev(A,\varep,u),\ev(B,s,v))_{C_\gamma} \mapsto P(\tou|\ev(A,\varep,u),\ev(C,\varep',u'),\ev(B,s,v))_\C.
\]
After averaging, we take the limit $\varep\to0$, and we obtain
\beqa
	\lefteqn{\forall \delta>0:\; \exists \varep_0>0 \;|\; \forall \varep'<\varep_0:}  && \n &&
	-\delta<\lim_{\varep\to0} \lt[P(\tou|\ev(A,\varep,u),\ev(B,s,v),\ev(C,\varep',u'))_\C-P(\tou|\ev(A,\varep,u),\ev(B,s,v))_C\rt]<\delta. \no
\eeqa
But $\lim_{\varep\to0} P(\tou|\ev(A,\varep,u),\ev(B,s,v),\ev(C,\varep',u'))_\C = P(\tou|\ev(B,s,v),\ev(C,\varep',u'))_{\C\setminus\b{A}}$ exists. Hence,
\beqa
	\lefteqn{\forall \delta>0:\; \exists \varep_0>0 \;|\; \forall \varep'<\varep_0:}  && \n &&
	-\delta<P(\tou|\ev(B,s,v),\ev(C,\varep',u'))_{\C\setminus\b{A}}-\lim_{\varep\to0}P(\tou|\ev(A,\varep,u),\ev(B,s,v))_C<\delta.\no
\eeqa
Since we can choose $\delta$ as small as we want, this shows that $\lim_{\varep\to0}P(\tou|\ev(A,\varep,u),\ev(B,s,v))_C$ exists, and that it is equal to the right-hand side of (\ref{eqpr2}).

Finally, in (\ref{eqpr2}) we make $B$ tend to $C$ and at the same time the partner of $B$ tend to $A$ (so that $\ev(B,s,v)$ is an increasing sequence of sets). The limit of the sequence of sets $!\ev(B,s,u)$ exists, and has measure zero in the measure induced by the limit $\varep\to0$ on both sides of (\ref{eqpr2}) thanks to assumption \ref{assborder}. Then, the event $\ev(B,s,v)_C$ has probability that tends to one on both sides, which proves (\ref{eqthcr2}).
\eproof

The theorem above is at the basis of the definition of a probability function on doubly connected domains. Since we have independence from the quantity $u$, we define it as follows:
\begin{defi}\label{nsc}
The probability of an event $\tou$ on $C\setminus \cl{A}$, for $A,\, C$ simply connected domains with $\cl{A}\subset C$, and $\supp(\tou)\subset C\setminus \cl{A}$, is defined by
\beq
    P(\tou)_{C\setminus\b{A}} \equiv \lim_{\varep\to0} P(\tou|\ev(A,\varep,u))_C~.
\eeq
\end{defi}
The fact that this is a correct definition is guaranteed by the following theorem, ensuring that conformal invariance holds for any transformations that are conformal on the annular region.
\begin{theorem}\label{thcr3}
Probabilities on $C\setminus \cl{A}$, for $A,\, C$ simply connected domains with $\cl{A}\subset C$, are invariant under conformal transformations $g:C\setminus\cl{A} \to C^\sharp\setminus\cl{A^\sharp}$ where $A^\sharp\subset \C$ and $C^\sharp\subset \C$ are simply connected domains with $\cl{A^\sharp} \subset C^\sharp$:
\beq
    P(g(\tou))_{C^\sharp\setminus\cl{A^\sharp}} = P(\tou)_{C\setminus \cl{A}}~.
\eeq
\end{theorem}
\proof Thanks to theorem \ref{thcr2}, the probability on $C\setminus \cl{A}$ is invariant under conformal transformations $C\to C'$ and conformal transformations $\C\setminus\cl{A} \to \C\setminus \cl{A'}$, with $C'$ and $A'$ simply connected domains. Any conformal transformation $g$ as in the theorem is a combination of such conformal transformations.
\eproof

Besides obtaining the probability function on doubly connected domains from that on simply connected domains, we will need to have the opposite: when one of the components of the boundary of an annular domain becomes very small, say, then in the limit we recover a simply connected domain.
\begin{theorem}\label{theopt}
Consider $A,\, C$ two simply connected domains with $\cl{A}\subset C$, two points $z\in A$, $z'\not\in C$, and an event $\tou$ supported in $C\setminus \cl{A}$. Then we have
\beq\label{ept}
	\blim_{\lambda\to0} P(\tou)_{C\setminus \lambda_{z,z'}\cl{A}} = P(\tou)_C.
\eeq
This holds also for $C=\C$.
\end{theorem}
\proof First, the fact that (\ref{ept}) holds for $C=\C$ is just a consequence of theorem \ref{thlimC}. For the case where $C$ is a simply connected domain, we re-write the probability on the left-hand side of (\ref{ept}), for fixed $\lambda$, using definition \ref{nsc}, as
\beq
	\lim_{\varep\to0} P(\tou|\ev(C,\varep,u))_{\C\setminus \lambda_{z,z'}\cl{A}} =
	\lim_{\varep\to0} P(\lambda_{z',z}\tou|\lambda_{z',z}\ev(C,\varep,u))_{\C\setminus \cl{A}}.
\eeq
From this, we can re-trace the lines of the proof of theorem \ref{thlimC} up to the equivalent of equation (\ref{eqfppr}), using  restricted re-randomisation, the existence of the limit $\varep\to0$ (theorem \ref{thcr2}), and the continuity of the limit of the restricted probability (theorem \ref{theores1}). That is,
\beq
	\lim_{\varep\to0} P(\lambda_{z',z}\tou|\lambda_{z',z})_{\C\setminus \cl{A}}
	=
	\mathbb{E}_{\gamma_\lambda^B} \lt[\lim_{\varep\to0}
		P(\lambda^{-1} \circ g\circ\lambda_{z',z} \tou|\lambda^{-1} \circ g\circ\lambda_{z',z}\ev(C,\varep,u)
		)_{\lambda^{-1}\uD_{\gamma_\lambda^B}}\rt]
\eeq
by restricted re-randomisation (\ref{relation4rest}), theorem \ref{thcr2}, and Lebesgue's bounded convergence theorem; and
\beq
	\lim_{\lambda\to0} \lim_{\varep\to0} P(\lambda_{z',z}\tou|\lambda_{z',z}\ev(C,\varep,u))_{\C\setminus \cl{A}}
	=
	\lt\bra \lim_{\varep\to0} P(h\tou |h \ev(C,\varep,u))_{\t{C}} \rt\ket_{ h(\supp(\tou)\cup\supp(\tou'))}
\eeq
by theorem \ref{theores1} (and corolary \ref{corolconj}). Finally, Lebesgue's bounded convergence theorem can be used again to write the right-hand side as
\[
	\lim_{\varep\to0}\bra P(h\tou |h \ev(C,\varep,u))_{\t{C}} \ket_{ h(\supp(\tou)\cup\supp(\tou'))}
\]
and by (\ref{formularest}), we obtain
\beq
	\lim_{\lambda\to0} \lim_{\varep\to0} P(\lambda_{z',z}\tou|\lambda_{z',z}\ev(C,\varep,u))_{\C\setminus \cl{A}} =
	\lim_{\varep\to0} P(\tou|\ev(C,\varep,u))_\C.
\eeq
Using theorem \ref{thcr1}, this completes the proof.
\eproof

Finally, we prove another result in the spirit of theorem \ref{thcr2}. This result again agrees with the intuition that it is the small loops near to $\p A$ that govern the vanishing of probabilities involving $\ev(A,\varep,u)$ as $\varep\to0$. The idea is essentially inspired from theorem \ref{thcr2}: we should be able to replace $\tou$ by an event of the type $\ev(B,\varep,u)$, in order to modify the domain of definition to $B$. However, this needs a careful analysis, since the interchange of the limits involved is non-trivial. One important conclusion of the following theorem is that the limit on the left-hand side of (\ref{eqcorss}) actually exists. This will be at the basis of the re-normalisation process defining the stress-energy tensor.

\begin{theorem}\label{corss}
For $A,\,B$ two simply connected domains with $A\subset B\subset C$, and $C$ a simply connected domain or $C=\C$, and with $\p A,\,\p B,\, \p C$ not intersecting each other, we have
\beq\label{eqcorss}
    \lim_{\varep\to 0} \frc{P(\ev(A,\varep,u))_B }{P(\ev(A,\varep,u))_{C}} =
    \lim_{\varep\to 0} \frc{P(\ev(B,\varep,u'))_{C\setminus \cl{A}} }{P(\ev(B,\varep,u'))_C}.
\eeq
In particular, both limits exist and are independent of $u$ and $u'$.
\end{theorem}
\proof Using theorem \ref{thcr1} we can write the left-hand side as
\[
    \lim_{\varep\to0}\lim_{\varep'\to0}\frc{P(\ev(A,\varep,u),\ev(B,\varep',u'))_C }{P(\ev(A,\varep,u))_{C} P(\ev(B,\varep',u'))_C}
\]
and, from definition \ref{nsc}, the right-hand side as
\[
    \lim_{\varep'\to0}\lim_{\varep\to0}\frc{P(\ev(A,\varep,u),\ev(B,\varep',u'))_C }{P(\ev(A,\varep,u))_{C} P(\ev(B,\varep',u'))_C}.
\]
In these expressions, the first limit taken always exists, thanks to theorems \ref{thcr1} and \ref{thcr2}.

Let us consider
\[
	\frc{P(\ev(A,\varep,u),\ev(B,\varep',u'))_C }{P(\ev(A,\varep,u))_{C} P(\ev(B,\varep',u'))_C}.
\]
Let us consider this as a probability restricted on $\ev(B,\varep',u')$. In the case where $C=\C$, we use (\ref{formularest}), where we may take $h=\id$ by appropriate choices of $A,B,C$ (without loss of generality by global conformal invariance). In order to cover both cases $C=\C$ and $C$ a simply connected domain at the same time, we will write $\t{C} = C$ if $C\neq \C$, and understand $\t{C}$ as the random domain in (\ref{formularest}) in the case $C=\C$. Then, from theorem \ref{theoCLEomega}, we may write the expression above as, with $\alpha = \p B$ and $\beta$ the boundary of the partner of $B$,
\[
	\int  \frc{P(\ev(A,\varep,u))_{B_\gamma}}{P(\ev(A,\varep,u))_{C}}\,d\omega_{\t{C},\alpha,\beta}(\gamma)
\]
for any $\varep'$ small enough. Since the limit $\varep\to0$ of this average exists, and since the function of $\gamma$ that is being averaged,
\beq\label{ratioBC}
	\frc{P(\ev(A,\varep,u))_{B_\gamma}}{P(\ev(A,\varep,u))_{C}},
\eeq
is strictly positive for all $\varep>0$, the limit $\varep\to0$ of this function is finite almost surely with respect to the measure on $\gamma$. This is true for any choice of $B$, of $\varep'$ and of $u'$. Let us write the limit $\varep\to0$ of the ratio (\ref{ratioBC}) as $(B_\gamma,C)$. For any $C$ and $C'$ with $B\subset C$ and $B\subset C'$, there is a $B_\gamma$ such that both $(B_\gamma,C)$ and $(B_\gamma,C')$ are finite, because the measures on $\gamma$ obtained from $C$ and $C'$ are absolutely continuous with respect to each other. Dividing both ratios, we obtain $(C',C)$, hence this ratio is finite for any $C$ and $C'$. Hence, the ratio (\ref{ratioBC}) is in fact finite for any $\gamma$ lying betwen $\alpha$ and $\beta$.

Then, we may wish to use theorem \ref{theores2}. In order to do so, however, we consider rather
\[
	\frc{P(\ev(A,\varep,u),\ev(B,\varep',u')|\ev(D,s,v))_{\t{C}} }{
		P(\ev(A,\varep,u)|\ev(D,s,v))_{C} P(\ev(B,\varep',u')|\ev(D,s,v))_{\t{C}}}.
\]
where $\p D$ lies between $\p A$ and $\p B$. We construct again the loop $\gamma$, so that we consider
\beq\label{exprBC}
	\frc{P(\ev(A,\varep,u)|\ev(D,s,v))_{B_\gamma}}{P(\ev(A,\varep,u)|\ev(D,s,v))_C}.
\eeq
By the same argument as above, the limit $\varep\to0$ is finite. We may further write this ratio as an average over a random curve $\t\gamma$ of
\beq\label{ratioDC}
	\frc{P(\ev(A,\varep,u))_{D_{\t\gamma}}}{P(\ev(A,\varep,u)|\ev(D,s,v))_C}.
\eeq
Writing this as
\[
	\frc{P(\ev(A,\varep,u))_{D_{\t\gamma}}}{P(\ev(A,\varep,u)|\ev(D,s,v))_C}
	= \frc{P(\ev(A,\varep,u))_{D_{\t\gamma}}}{P(\ev(A,\varep,u))_C}
		\frc{P(\ev(D,s,v))_{C}}{P(\ev(D,s,v)|\ev(A,\varep,u))_C}
\]
we see that the limit of the second factor exists thanks to theorem \ref{thcr1} (for $C=\C$) or \ref{thcr2} (for $C$ a simply connected domain), and that the limit of the first factor is finite thanks to the considerations above. Hence, the limit $\varep\to0$ of the ratio (\ref{ratioDC}) is finite. Hence, for the use of theorem \ref{theores2}, we take $a\mapsto \varep$, $c_a \mapsto 1/P(\ev(A,\varep,u)|\ev(D,s,v))_C$ and $\tou_a\mapsto \ev(A,\varep,u)$. Then, the theorem shows that the expression (\ref{exprBC}) is continuous as a function of $B_\gamma$, with a continuity statement that is independent of $\varep$.

Using $B_\gamma\to B$ as $\varep'\to0$, we may then simply repeat the steps of the proof of theorem \ref{thcr2}. First,
\beqa
	\lefteqn{\forall \delta>0:\; \exists \varep_0>0 \;|\; \forall \varep>0,\,\forall \varep'<\varep_0:}  && \n &&
	-\delta<\frc{P(\ev(A,\varep,u)|\ev(D,s,v))_{B_\gamma}}{P(\ev(A,\varep,u)|\ev(D,s,v))_C}-
		\frc{P(\ev(A,\varep,u)|\ev(D,s,v))_{B}}{P(\ev(A,\varep,u)|\ev(D,s,v))_C}<\delta. \no
\eeqa
Averaging over $B_\gamma$ (and $\t{C}$ if required), we have
\[
	\frc{P(\ev(A,\varep,u)|\ev(D,s,v))_{B_\gamma}}{P(\ev(A,\varep,u)|\ev(D,s,v))_C} \mapsto
	\frc{P(\ev(A,\varep,u),\ev(B,\varep',u')|\ev(D,s,v))_C }{P(\ev(A,\varep,u)|\ev(D,s,v))_{C} P(\ev(B,\varep',u')|\ev(D,s,v))_C}.
\]
We can then take the limit $\varep\to0$, and we obtain
\beqa
	\lefteqn{\forall \delta>0:\; \exists \varep_0>0 \;|\; \forall \varep'<\varep_0:}  && \n &&
	-\delta<
	\frc{P(\ev(B,\varep',u')|\ev(D,s,v))_{C\bs \overline{A}} }{P(\ev(B,\varep',u')|\ev(D,s,v))_C}
	- \lim_{\varep\to0} \frc{P(\ev(A,\varep,u)|\ev(D,s,v))_{B}}{P(\ev(A,\varep,u)|\ev(D,s,v))_C}
	<\delta. \no
\eeqa
Since we can choose $\delta$ as small as we want, this shows
\[
	\lim_{\varep\to0}\frc{P(\ev(A,\varep,u)|\ev(D,s,v))_{B}}{P(\ev(A,\varep,u)|\ev(D,s,v))_C}
	= \lim_{\varep\to0} \frc{P(\ev(B,\varep,u')|\ev(D,s,v))_{C\bs \overline{A}}}{P(\ev(B,\varep,u')|\ev(D,s,v))_C}
\]
In order to use assumption \ref{assborder}, we need to re-write the factors in such a way that we have explicitly an appropriate induced measure where the event $\ev(D,s,v)$ is evaluated. We have for instance
\[
	P(\ev(A,\varep,u)|\ev(D,s,v))_B= P(\ev(D,s,v)|\ev(A,\varep,u))_B \frc{P(\ev(A,\varep,u))_B}{P(\ev(D,s,v))_B}
\]
and similarly for the other factors. Using theorem \ref{thcr1} and definition \ref{nsc}, we then find
\beqa
	\lefteqn{\frc{P(\ev(D,s,v))_C P(\ev(D,s,v))_{B\bs\overline{A}}}{P(\ev(D,s,v))_B P(\ev(D,s,v))_{C\bs\overline{A}}}
	\lim_{\varep\to0}\frc{P(\ev(A,\varep,u))_{B}}{P(\ev(A,\varep,u))_C}} \no && \\
	&=& \frc{P(\ev(D,s,v))_C \t{P}(\ev(D,s,v))_{B\bs\overline{A}}}{P(\ev(D,s,v))_{C\bs\overline{A}} P(\ev(D,s,v))_{B}}
	\lim_{\varep\to0} \frc{P(\ev(B,\varep,u'))_{C\bs \overline{A}}}{P(\ev(B,\varep,u'))_C}
	\label{icorss}
\eeqa
where we define
\[
	\t{P}(\ev(D,s,v))_{B\bs\overline{A}} = \lim_{\varep\to0} P(\ev(D,s,v)|\ev(B,\varep,u'))_{C\bs\overline{A}}.
\]
Naturally, we expect that $\t{P}(\ev(D,s,v))_{B\bs\overline{A}} = P(\ev(D,s,v))_{B\bs\overline{A}}$, but we have not proved it yet. Here, we only need to bring $D$ towards $B$ and its partner towards $A$ (so that $\ev(D,s,v)$ is an increasing sequence of sets), and use assumption \ref{assborder} in combination with theorem \ref{thcr2}, in order to have $\t{P}(\ev(D,s,v))_{B\bs\overline{A}} = P(\ev(D,s,v))_{B\bs\overline{A}}=1$; we obtain (\ref{eqcorss}). Clearly, from (\ref{icorss}), the ratio $\t{P}(\ev(D,s,v))_{B\bs\overline{A}}/P(\ev(D,s,v))_{B\bs\overline{A}}$ is independent of $D,\,s,\,v$, hence should be 1, so that we also have proved the equality between these two factors. \eproof

\sect{Conclusion}

In the present paper, the first part of a work concerning the construction of the stress-energy tensor in CLE, we presented an introduction to CLE, we defined notions of continuity and of support of CLE events, we introduced definitions for probability functions on $\C$ and on annular domains, and we proved certain theorems concerning all these notions. In particular, we proved that any event is continuous at a domain of definition containing its support. This, and other more precise and slightly more technical continuity theorems, allowed us to prove conformal invariance and other technical theorems for the probability functions on $\C$ and on annular domains (up to three assumptions). The most important theorems, at the basis of the stress-energy tensor construction and of the conformal Ward identities, are theorems \ref{thcr2}, \ref{thcr3} and \ref{corss}. They will form the starting point of the second part of this work. The construction of a conformally invariant probability function on $\C$, definition \ref{defCplane} and theorem \ref{thglobinv}, also plays a fundamental r\^ole in the second part, being at the heart of our derivation of the Schwarzian derivative term in the transformation property of the stress-energy tensor.

\appendix

\sect{Some notations and conventions} \label{notations}

In this appendix we present some notations and conventions used in this paper.

The notation $\C$ will be understood as representing the Riemann sphere, the set of complex numbers $+\{\infty\}$ with the sphere topology. We take domains as open sets in $\C$, and futher restrict to Jordan domains. For $A$ a domain, $\cl{A}$ is its closure. The domain $\uD$ is the disk of unit radius centered around the point 0 in $\C$, $\uD = \{z\in \C\,|\,|z|<1\}$. Following the usual nomenclature in CFT, we term ``global conformal transformation'' any transformation that is conformal on $\C$ (that is, that preserves angles everywhere in $\C$.)

We need a generalised ``scale transformation'', which we denote by $\lambda_{z_1,z_2}$ for $\lambda\in\R^+$ and $z_1,z_2\in \C$. It is defined by
\beq\label{genscale}
    \lambda_{z_1,z_2} (x) = g^{-1}(\lambda g(x))~,\quad g(x) = \frc{x-z_1}{x-z_2}
\eeq
giving
\beq
	\lambda_{z_1,z_2}(x) = \frc{(1-\lambda)z_1z_2 - (z_1-\lambda z_2) x}{z_2-\lambda z_1 - (1-\lambda) x}.
\eeq
The conformal transformation $g$ sends $z_1$ to $0$ and $z_2$ to $\infty$. Hence, $\lambda_{z_1,z_2}$ for $\lambda$ increasing represents a flow from the point $z_1$ to the point $z_2$, which are two fixed points. The usual scale transformation is the case $\lambda_{0,\infty}$. Note that the function $g$ can be rescaled and rotated, $g\mapsto \lambda' g$ for $\lambda'\in\C,\,0<|\lambda'|<\infty$, without affecting $\lambda_{z_1,z_2}$. Hence $g$ can be taken as any global conformal transformation that takes $z_1$ to $0$ and $z_2$ to $\infty$. Particular cases are
\beq
	\lambda_{z_1,\infty}(x) = z_1+\lambda(x-z_1),\quad \lambda_{\infty,z_2}(x) = z_2 + \lambda^{-1} (x-z_2)
\eeq
and we have
\beq
	\lambda_{z_1,z_2}(x) = z_1 +\lambda (z_1-z_2) \frc{x-z_1}{x-z_2} + O(\lambda^2)
\eeq
for $z_1\neq \infty$.

It will be convenient to have some concept of the {\em extent} of a loop, or of any set, in a simply connected domain $C\subset\C$. For every simply connected domain $C$, let us fix a conformal mapping $C\to\uD$. We will take the extent to be the minimal diameter of a closed disk in $\uD$ that completely contains the mapped loop or set. That is, with $g_C:C\to\uD$ and $d$ the usual $\R^2$ distance function, the extent of a set $A$ lying in $C$ is ${\rm sup}_{x_1\in g_C(A),\,x_2\in g_C(A)} d(x_1,x_2)$. This definition depends on our choice of conformal mappings, and on the domain $C$ where the set lies, but we will only need its general properties. Likewise, we define the {\em radius} of a set to be half its diameter, and the {\em distance between two points} to be the extent of the two-point set.

We find it useful to introduce the following conventions concerning sets of loops, etc.:
\begin{enumerate}
\item \label{pointconfC} $\conf_C$: the set of all CLE configurations (see subsection \ref{subsectaxioms}) where all loops are
contained in $C$, with $C$ a simply connected domain, or $C=\C$.

\item \label{pointev} $\evs(\conf_C)$: the set of all events on $C$, that is, of all subsets of $\conf_C$.

\item $\tou_C$: for $\tou\in\evs(\conf_{\C})$ an event and $C$ a simply connected domain, this is the restriction of $\tou$ on the $\conf_C$:
\beq\label{touC}
    \tou_C = \tou \cap \conf_C \in \evs(\conf_C).
\eeq

\item $\pres$: The characteristic function,
\beqa
    \pres \;:\; \conf_\C \otimes \evs(\conf_\C) &\to& \{0,1\}\n
    x \otimes \tou &\mapsto& \pres(x,\tou) = \lt\{\ba{ll} 1 & (x\in\tou) \\ 0 &
    (x\not\in\tou) \ea\rt. \label{pres}
\eeqa

\item $P_{\sam}$: the probability function associated to a sequence $\sam = (x_1,x_2,\ldots)$ of configurations,
\beqa
    P_{\sam}\; :\; \evs(\conf_\C) &\to& [0,1] \n
    \tou &\mapsto& P_\sam(\tou) = \lim_{N\to\infty}
    N^{-1}\sum_{j=1}^N \pres(x_j,\tou). \label{Psam}
\eeqa

\item $C_\gamma$: for $C$ a simply connected domain and $\gamma\subset C$ a simple loop, this is the simply connected domain bounded by the loop $\gamma$ and not containing $\p C$; that is, the interior of $\gamma$ in $C$.

\item $\inter_C(x)$: for $x\in\conf_{\C}$ a configuration on $\C$, and $C$ a domain, this is the set of loops in $x$ completely contained inside $C$:
\beq
    \inter_C(x) = \{\gamma\in x\,|\,\gamma\subset C\} \in \conf_C
\eeq

\item $\exter_C(x)$: for $x\in\conf_{\C}$ a configuration on $\C$, and $C$ a domain, this is the set of loops in $x$ completely contained outside $C$:
\beq
    \exter_C(x) = \{\gamma\in x\,|\,\gamma\subset \C\setminus \cl{C}\} \in \conf_{\C\setminus \cl{C}}
\eeq

\end{enumerate}

\end{document}